\documentstyle[11pt,a4wide]{article}

\begin{document}

\begin{titlepage}
\noindent
\begin{flushright}
PRA -- HEP 98/4\\
March 1998\\
\end{flushright}

\vfill

\begin{center}
\noindent
{\huge\bf{Point-splitting regularization
of composite operators and anomalies}}

\vspace{1cm}

{\large J. Novotn\'{y}{\footnote{e-mail: {\tt
Jiri.Novotny@mff.cuni.cz}}} and M. Schnabl{\footnote{e-mail:
{\tt schnabl@hp01.troja.mff.cuni.cz}}} \\}
{\it Nuclear Centre, Faculty of Mathematics and Physics, Charles University\\
V Hole\v{s}ovi\v{c}k\'{a}ch 2, 180 00 Prague 8, Czech Republic\\}

\end{center}

\vfill

\begin{center}
{\bf Abstract}
\end{center}

The point-splitting regularization technique for composite operators is
discussed in connection with anomaly calculation.
We present a pedagogical
and self-contained review of the topic with an emphasis
on the technical details.
We also develop  simple algebraic tools to handle the path ordered
exponential insertions used within the covariant and non-covariant
version of the point-splitting method.
The method is then applied to the calculation of the chiral, vector,
trace, translation and Lorentz anomalies within diverse versions
of the point-splitting regularization and a connection between the
results is described.
As an alternative to the standard approach we use the idea of deformed
point-split transformation and corresponding Ward-Takahashi identities
rather than an application of the equation of motion, which seems to
save the complexity of the calculations.


\vfill

\end{titlepage}


\section{Introduction}


The point-splitting regularization of the composite operators is a method
which has a long history. The early works \cite{history} date back more than
fifty years ago. The idea of the method is very simple: the slightly
different space time points are assigned to the elementary fields from which
a composite operator is built. As a result, the short distance singularities
of the composite operator which appear in the limit of the coinciding points
are regulated. E.g. the fermionic electromagnetic current
\begin{equation}
J_\mu ^{em}(x)=e\overline{\psi }(x)\gamma _\mu \psi (x)
\end{equation}
can be regulated according to the prescription
\begin{equation}
J_\mu ^{em}(x)^{{\rm reg}}=e\overline{\psi }(x+\varepsilon )\gamma _\mu \psi
(x-\varepsilon )  \label{naive}
\end{equation}
and the short distance (UV) singularities can be isolated as the divergent
terms in the expansion of the regularized expression for $\varepsilon
\rightarrow 0.$ However, the naive form of the point-splitting (\ref{naive})
suffers from the violation of the $U(1)_{em}$ gauge invariance. This
loophole can be cured introducing a suitable compensating factor, the gauge
invariant regulated current can be constructed as
\begin{equation}
J_\mu ^{em}(x)^{{\rm reg}}=e\overline{\psi }(x+\varepsilon )\gamma _\mu \psi
(x-\varepsilon )\exp \left( {\rm i}e\int_{x-\varepsilon }^{x+\varepsilon }%
{\rm d}y\cdot A(y)\right) .  \label{invar}
\end{equation}
Various modifications of this approach were described in the literature \cite
{modification}, and the point-splitting method still attracts interest in
various contexts \cite{Osland-Wu}.

One of the most important applications of this regularization scheme is the
calculation of the quantum anomalies. The first calculation of such a type
was done already in the pioneering work \cite{Jackiw} on the
Adler-Bell-Jackiw anomaly and was also systematically used in the
fundamental work \cite{Bardeen} to get the general form of the nonabelian
anomaly. It also proved to be a suitable tool for the calculation of the
anomalies connected with the trace and divergence of the energy-momentum
tensor \cite{energy-momentum} in the gravitational background. Since the
time of the first works on this topic, the point-splitting regularization
method has become a well-understood standard routine.

In this short review we attempt to present a self-contained pedagogical
introduction to the method and its application to the anomaly calculation
with an emphasis on the technical details and the language of the
contemporary field theory. We also offer a rather nonstandard approach to
the derivation of the anomalies in terms of the Ward-Takahashi identities
for deformed point-split transformations rather then using the equations of
motion, which seems to save the complexity of the calculations.

The paper is organized as follows. In Section 2 we concentrate on the
problem of the short distance singularities of the fermion propagator in the
background of a nonabelian gauge field and the heat kernel method is briefly
reviewed. In Section 3 we outline the general strategy of the anomalies
calculation. In Section 4, 5, 6, 7 and 8 we illustrate this general strategy
using the concrete examples of the chiral, trace, translational and Lorentz
anomalies respectively. The non-covariant point-splitting and the
corresponding modifications of the anomalies are described in Section 9. The
properties of the vector current in various versions of the point-splitting
and integrability of the vector current are discussed in Section 10. Some of
the technical details are postponed to Appendices A and B.



\section{The singularities of the fermion propagator at short distances, the
heat kernel method\label{asymptotic}}

In this section we give a pedagogical overview of one of the most efficient
methods for obtaining the short distance properties of the euclidean
propagator of the fermions in the background nonabelian gauge field. This
method is based on the properties of the so called heat kernel, associated
with a suitable elliptic differential operator of the second order. We also
introduce some notation which will be useful in the rest of the article.

Let us first consider the following elliptic operator, operating on the
sections of some Hermitian vector bundle over four-dimensional flat
Euclidean space

\begin{equation}
\Delta =-D^2+V,  \label{operator}
\end{equation}
where the covariant derivative is given as

\begin{equation}
D=\partial +A
\end{equation}
and the gauge field $A$ and the (positive) potential $V$ satisfy the
following hermiticity properties
\begin{equation}
A^{+}=-A,\;V^{+}=V,
\end{equation}
i.e. the operator $\Delta $ is positive and Hermitian. Let us define an
Euclidean ``scalar propagator with mass $m$'' as the $x$-representation
matrix element of the inverse of the operator $\Delta +m^2$

\begin{equation}
G(x,y;m^2)=(x|(\Delta +m^2)^{-1}|y).
\end{equation}
First we will investigate the $|x-y|\rightarrow 0$ asymptotics of this
propagator with a general potential $V$ . The result (with specific choice
of the potential $V$ ) will be then used to obtain the short distance
behavior of the fermion propagator defined as the inverse of the Dirac
operator $\gamma \cdot D+{\rm i}m$

\begin{equation}
S(x,y;m)=(x|(\gamma \cdot D+{\rm i}m)^{-1}|y)
\end{equation}
where $\gamma _\mu $ are the antihermitian Euclidean $\gamma -$ matrices,
which satisfy the anticommutation relations
\begin{equation}
\{\gamma _\alpha ,\gamma _\beta \}=-2\delta _{\alpha \beta }.
\end{equation}

Let us introduce now the key object of our further considerations. The heat
kernel ${\cal G}(x,y;\tau )$ of the elliptic operator $\Delta $ given by the
formula (\ref{operator}) is defined as the kernel (i.e. the $x$%
-representation matrix element) of the operator ${\rm e}^{-\tau \Delta }$,
explicitly

\begin{equation}
{\cal G}(x,y;\tau )=(x|{\rm e}^{-\tau \Delta }|y).
\end{equation}
The heat kernel satisfies the following partial differential equation,

\begin{eqnarray}
-\frac \partial {\partial \tau }{\cal G}(x,y;\tau )=\Delta {\cal G}(x,y;\tau
),  \label{equation}
\end{eqnarray}
with the initial condition
\begin{equation}
\lim\limits_{\tau \rightarrow 0}{\cal G}(x,y;\tau )=\delta ^{(4)}(x-y).
\label{initial condition}
\end{equation}
Let us summarize some of its properties. If we find the solution of the
equation (\ref{equation}) in the form \footnote{%
Let us note, that $\Delta _E^{(4)}(x,0)=\int_0^\infty {\rm d}\tau \frac
1{(4\pi \tau )^2}{\rm e}^{-\frac{|x-y|^2}{4\tau }}=\frac 1{4\pi ^2}\frac
1{|x-y|^2}$ corresponds to the free scalar Euclidean propagator with zero
mass.}

\begin{eqnarray}
{\cal G}(x,y;\tau )=\frac 1{(4\pi \tau )^2}{\rm e}^{-\frac{|x-y|^2}{4\tau }%
}F(x,y;\tau ),
\end{eqnarray}
the initial condition (\ref{initial condition}) requires then

\begin{equation}
\lim\limits_{\tau \rightarrow 0}F(x,y;\tau )=1.
\end{equation}
Assuming for the function $F(x,y;\tau )$ the following asymptotic expansion
\cite{Gilkey} for $\tau \rightarrow 0$

\begin{equation}
F(x,y;\tau )=\sum_{n=0}^\infty a_n(x,y)\tau ^n,  \label{asymptotics}
\end{equation}
the coefficients of this expansion satisfy then the following set of
recursion relations, which can be obtained by inserting the asymptotics (\ref
{asymptotics}) to the equation (\ref{equation}) with the initial condition (%
\ref{initial condition})
\begin{eqnarray}
(x-y)\cdot D_xa_0(x,y) &=&0,\;a_0(x,x)=1,  \nonumber \\
na_n(x,y)+(x-y)\cdot D_xa_n(x,y) &=&-\Delta _xa_{n-1}(x,y).  \label{Seeley}
\end{eqnarray}
The coefficients $a_n(x,y)$ (known as the Seeley-DeWitt coefficients%
\footnote{%
For a review and complete list of references see also \cite{Ball}.} \cite
{Seeley-deWitt}) are smooth for $|x-y|\rightarrow 0$. Let us indicate, how
these relations can be explicitly solved. Introducing (with $x$ and $y$
fixed)
\begin{equation}
x_t=y+t(x-y)
\end{equation}
and using the first relation (\ref{Seeley}), we get for $a_0(x_t,y)$ the
following ordinary differential equation

\begin{equation}
\frac{{\rm d}}{{\rm d}t}a_0(x_t,y)=-(x-y)\cdot A(x_t)a_0(x_t,y),
\end{equation}
with the initial condition $a_0(x_0,y)=1$, which can be easily solved in
terms of the $T$ -ordered exponential

\begin{equation}
a_0(x_t,y)=T\exp \left( -\int_0^t{\rm d}t(x-y)\cdot A(x_t)\right) .
\end{equation}
$a_0(x,y)$ is then given by the following formula

\begin{equation}
a_0(x,y)=a_0(x_1,y)=T\exp \left( -\int_0^1{\rm d}t(x-y)\cdot A(x_t)\right)
=\Omega (x,y)
\end{equation}
as the parallel transporter $\Omega (x,y)$ along the straight line
connecting the points $x$ and $y$ $.$ In the same way we get the following
differential equation for $a_1(x_t,y)$

\begin{equation}
\frac{{\rm d}}{{\rm d}t}ta_1(x_t,y)=-(x-y)\cdot A(x_t)ta_1(x_t,y)-\Delta
_{x_t}a_0(x_t,y),
\end{equation}
the solution of which can be expressed in terms of the known coefficient $%
a_0(x_t,y)$%
\begin{equation}
ta_1(x_t,y)=-a_0(x_t,y)\int_0^t{\rm d}\tau a_0(x_\tau ,y)^{-1}\Delta
_{x_\tau }a_0(x_\tau ,y),
\end{equation}
therefore

\begin{eqnarray}
a_1(x,y) &=&-a_0(x,y)\int_0^1{\rm d}ta_0(x_t,y)^{-1}\Delta _{x_t}a_0(x_t,y)
\nonumber \\
\ &=&-\int_0^1{\rm d}ta_0(x,x_t)\Delta _{x_t}a_0(x_t,y)
\end{eqnarray}
(here we used simple properties of the coefficient $a_0(x,y),$ namely $%
a_0(x,y)^{-1}=a_0(y,x)$ and $a_0(x,y)a_0(y,z)=a_0(x,z);$ valid for $x,y,z$
on the same line). The generalization of this procedure for general $n$ is
straightforward, the general formula reads
\begin{equation}
a_n(x,y)=-\int_0^1{\rm d}tt^{n-1}a_0(x,x_t)\Delta _{x_t}a_{n-1}(x_t,y).
\end{equation}

Let us now use the properties of the heat kernel to get the short distance
asymptotics of the scalar propagator $G(x,y;m^2)$. This can be achieved by
expressing it in terms of the heat kernel as
\begin{equation}
G(x,y;m^2)=\int_0^\infty {\rm d}\tau {\rm e}^{-\tau m^2}{\cal G}(x,y;\tau );
\label{integral representation}
\end{equation}
this formula is known as the Schwinger proper-time representation of the
propagator. The short distance asymptotics of $G(x,y;m^2)\,$ can be then
obtained as a consequence of the $\tau \rightarrow 0$ asymptotics of the
heat kernel ${\cal G}(x,y;\tau ).$ Indeed, from the Schwinger representation
of the scalar propagator (\ref{integral representation}) we have
\begin{equation}
G(x,y;m^2)=\int_0^\infty \frac{{\rm d}\tau }{(4\pi \tau )^2}{\rm e}^{-\tau
m^2-\frac{|x-y|^2}{4\tau }}F(x,y,\tau ),
\end{equation}
i.e. for $|x-y|\neq 0$ the integrand is well-behaved\footnote{%
Let us recall the limit $F(x,y,\tau )\rightarrow 1$ for $\tau \rightarrow 0$
and, in the finite volume, the asymptotic behaviour of the heat kernel $%
{\cal G}(x,y,\tau )\approx \exp (-\lambda _0\tau )$ $\Phi _0^{*}(y)$ $\Phi
_0(x)$ for $\tau \rightarrow \infty $, where $\lambda _0>0$ is the lowest
eigenvalue and $\Phi _0(x)$ the corresponding eigenvector of the operator $%
\Delta $.} both for $\tau \rightarrow 0$ and $\tau \rightarrow \infty $. For
$|x-y|=0$ we have however a nonintegrable singularity at $\tau \rightarrow 0$%
. In order to isolate the short distance singularities, let us divide the
integrand into two parts by means of adding and subtracting the leading
asymptotics of the integrand for $\tau \rightarrow 0$:

\begin{eqnarray}
G(x,y;m^2) &=&\int_0^\infty \frac{{\rm d}\tau }{(4\pi \tau )^2}{\rm e}%
^{-\tau m^2-\frac{|x-y|^2}{4\tau }}\sum_{i=0}^na_i(x,y)\tau ^i  \nonumber \\
&&\ \ +\int_0^\infty \frac{{\rm d}\tau }{(4\pi \tau )^2}{\rm e}^{-\tau m^2-%
\frac{|x-y|^2}{4\tau }}(F(x,y;\tau )-\sum_{i=0}^na_i(x,y)\tau ^i).
\end{eqnarray}

The second term on the right hand side has smooth limit for $%
|x-y|\rightarrow 0$ , if we take $n$ large enough, (the corresponding
integrand behaves as $\tau ^{n-1}$ for $\tau \rightarrow 0$ and $\tau ^{n-2}%
{\rm e}^{-\tau m^2}$ for $\tau \rightarrow \infty $). The source of the
singular behavior of the scalar propagator for $|x-y|\rightarrow 0$
corresponds to the first term.

Let us describe this in more detail. We have
\begin{equation}
G(x,y;m^2)=\sum_{i=0}^na_i(x,y)I_n(x-y)+R(x,y),  \label{singular part}
\end{equation}
where $R(x,y)$ is finite in the limit $\mid x-y$ $\mid \rightarrow 0$ and $%
I_n(x)$ are the following integrals
\begin{equation}
I_n(x)=\int_0^\infty \frac{{\rm d}\tau }{(4\pi \tau )^2}\tau ^n{\rm e}%
^{-\tau m^2-\frac{|x|^2}{4\tau }}=\frac{m^{-n+1}}{8\pi ^2}\left( \frac{|x|}%
2\right) ^{n-1}K_{-n+1}(m|x|).
\end{equation}
Here $K_n(z)$ are the MacDonald functions\footnote{%
Here
\par
\[
\psi (k+1)=-\gamma _E+\sum_{j=1}^k\frac 1j.
\]
}

\begin{eqnarray}
K_n(z) &=&\frac 12\sum_{k=0}^{n-1}\frac{(-1)^k(n-k-1)!}{k!}\left( \frac
z2\right) ^{2k-n}  \nonumber \\
&&+\frac{(-1)^{n+1}}2\sum_{k=0}^\infty \frac 1{k!(n+k)!}\left( \frac
z2\right) ^{2k+n}(2{\rm \ln }\left( \frac z2\right) -\psi (k+1)-\psi
(k+n+1)),
\end{eqnarray}
satisfying

\begin{equation}
K_n(z)=K_{-n}(z).
\end{equation}
Explicitly

\begin{eqnarray}
I_0(x) &=&\Delta _E^{(4)}(x;m^2)=\frac 1{4\pi ^2x^2}m|x|K_1(m|x|)  \nonumber
\\
&=&\frac 1{4\pi ^2}\left( \frac 1{x^2}+\frac{m^2}4{\rm \ln }m^2|x|^2-\frac{%
m^2}4(2{\rm \ln }2+2\gamma _E-1)\right)  \nonumber \\
&&+{\cal O}(m^2|x|^2,m^2|x|^2{\rm \ln }m^2|x|^2)
\end{eqnarray}
and

\begin{eqnarray}
I_1(x) &=&\Delta _E^{(2)}(x;m^2)=\frac 1{8\pi ^2}K_0(m|x|)  \nonumber \\
&=&-\frac 1{16\pi ^2}({\rm \ln }m^2|x|^2-2{\rm \ln }2+2\gamma _E-2)+{\cal O}%
(m^2|x|^2,m^2|x|^2{\rm \ln }m^2|x|^2)
\end{eqnarray}
where $\Delta _E^{(d)}(x-y;m^2)$ is the free Euclidean scalar propagator
(i.e. for $A=0$ and $V=0$) in $d$ dimensional euclidean space. Note also,
that for $n\geq 2$ , the $I_n(x)$ (and also $\partial _\mu I_n(x)$) are
regular for $x\rightarrow 0$, i.e. we can keep only the first two terms in
the sum on the right hand side of (\ref{singular part}) in order to pick up
explicitly the potentially singular part.

Therefore we have

\begin{equation}
G(x,y;m^2)=\frac 1{4\pi ^2}\left( \frac{a_0(x,y)}{|x-y|^2}+\frac 14{\rm \ln }%
m^2|x-y|^2(a_0(x,y)m^2-a_1(x,y))\right) +{\cal O}(1).
\label{scalar propagator}
\end{equation}

The fermion propagator, defined as

\begin{equation}
S(x,y;m)=(x|\frac 1{\gamma \cdot D+{\rm i}m}|y),
\end{equation}
can be rewritten in the form

\begin{eqnarray}
S(x,y;m) &=&(x|(\gamma \cdot D-{\rm i}m)\frac 1{(\gamma \cdot D)^2+m^2}|y)
\nonumber \\
&=&(\gamma \cdot D-{\rm i}m)(x|\frac 1{-D^2-\frac{{\rm i}}2\sigma _{\mu \nu
}F_{\mu \nu }+m^2}|y)  \nonumber \\
&=&(\gamma \cdot D-{\rm i}m)(x|\frac 1{\Delta +m^2}|y)
\end{eqnarray}
with positive operator

\begin{equation}
\Delta =-D^2-\frac{{\rm i}}2\sigma _{\mu \nu }F_{\mu \nu }.
\label{fermion operator}
\end{equation}
The matrices $\sigma _{\alpha \beta }=\frac{{\rm i}}2[\gamma _\alpha ,\gamma
_\beta ]$ are the generators of the $SO(4)$ euclidean rotations and
\begin{equation}
F_{\mu \nu }=[D_\mu ,D_\nu ]=\partial _\mu A_\nu -\partial _\nu A_\mu
+[A_\mu ,A_\nu ]
\end{equation}
is the gauge field strength. I.e., expressing the fermion propagator in
terms of the ``scalar propagator with mass $m$'' corresponding to the
operator (\ref{fermion operator}), we have for the potentially singular part
of the fermion propagator

\begin{eqnarray}
S(x,y;m) &=&(\gamma \cdot D_x-{\rm i}m)G(x,y;m^2)  \nonumber \\
&=&\frac 1{4\pi ^2}\left(\frac{(\gamma \cdot D-{\rm i}m)a_0(x,y)}{|x-y|^2}-\frac{%
2\gamma _\mu (x-y)_\mu a_0(x,y)}{|x-y|^4}
\right.
\nonumber \\
&&+\frac 14{\rm \ln }m^2|x-y|^2(\gamma \cdot D-{\rm i}%
m)(a_0(x,y)m^2-a_1(x,y))
\nonumber \\
&&
\left.
+\frac 12\frac{\gamma _\mu (x-y)_\mu }{|x-y|^2}((a_0(x,y)m^2-a_1(x,y))+%
{\cal O}(1)\right) .  \label{fermion propagator}
\end{eqnarray}

To isolate the singular part of the propagators (\ref{scalar propagator})
and (\ref{fermion propagator}), it remains now to expand the Seeley- de Witt
coefficients in power series in $\mid x-y$ $\mid .$ Before doing this, we
add one important note. The above defined scalar and fermion propagators
have one inconvenient property - they are not gauge covariant. E.g. for the
fermion propagator we have the following transformation relation with
respect to the gauge transformation $U(x)$:

\begin{equation}
S(x,y;m)\rightarrow U(x)S(x,y;m)U^{+}(y).
\end{equation}
As a consequence of this, if we introduce
\begin{equation}
\overline{x}=\frac 12(x+y),\;\varepsilon =\frac 12(y-x),
\end{equation}
the coefficients of the expansion of the propagators $G(x,y)$ and $S(x,y;m)$
in powers and logarithms of $\varepsilon $ are not gauge covariant functions
of $\overline{x}$.

For the calculations presented in the following sections we need rather
objects, which transform covariantly. For this purpose, let us note, that
the parallel transporter along the straight line (which is nothing but the
Seeley-de Witt coefficient $a_0(x,y)$)
\begin{equation}
\Omega (x,y)=P\exp \left( -\int_0^1{\rm d}t(x-y)\cdot A(y+t(x-y)\right)
=a_0(x,y).
\end{equation}
transforms according to the prescription
\begin{equation}
\Omega (x,y)\rightarrow U(x)\Omega (x,y)U^{+}(y).
\end{equation}
Let us therefore define the ``covariant propagator''
\begin{equation}
S^{{\rm cov}}(x,\varepsilon )=\Omega (x,x-\varepsilon )S(x-\varepsilon
,x+\varepsilon ;m)\Omega (x+\varepsilon ,x).
\end{equation}
The transformation properties of the gauge exponential ensure the following
transformation of the covariant propagator:
\begin{equation}
S^{{\rm cov}}(x,\varepsilon )\rightarrow U(x)S^{{\rm cov}}(x,\varepsilon
)U^{+}(x).
\end{equation}
In the same way we can ``improve'' the scalar propagator, which transforms
noncovariantly

\begin{equation}
G(x,y;m^2)\rightarrow U(x)G(x,y;m^2)U^{+}(y),
\end{equation}
and define the covariant propagator

\begin{equation}
G^{{\rm cov}}(x,\varepsilon )=\Omega (x,x-\varepsilon )G(x-\varepsilon
,x+\varepsilon ;m^2)\Omega (x+\varepsilon ,x)
\end{equation}
with covariant transformation law
\begin{equation}
G^{{\rm cov}}(x,\varepsilon )\rightarrow U(x)G^{{\rm cov}}(x,\varepsilon
)U^{+}(x).
\end{equation}
The main advantage of such covariant propagators is, that the $x-$dependent
coefficients of the expansion in the powers and logarithms of $\varepsilon $
are now gauge covariant functions.

From the expression (\ref{scalar propagator}) we get then

\begin{equation}
G^{{\rm cov}}(x,\varepsilon )=\frac 1{4\pi ^2}\left( \frac 1{|\varepsilon
|^2}+\frac 14{\rm \ln }m^2|\varepsilon |^2(m^2-a_0(x,x-\varepsilon
)a_1(x-\varepsilon ,x+\varepsilon )a_0(x+\varepsilon ,x))\right) +{\cal O}%
(1).
\end{equation}
To make use of this formula, we need the expansion of the expression $%
a_0(x,x-\varepsilon )a_1(x-\varepsilon ,x+\varepsilon )a_0(x+\varepsilon ,x)$
in the powers of $\varepsilon .$ We only quote the result of the calculation
here, the details can be found in Appendix A:
\begin{equation}
a_0(x,x-\varepsilon )a_1(x-\varepsilon ,x+\varepsilon )a_0(x+\varepsilon ,x)=%
\frac{{\rm i}}2\sigma _{\alpha \beta }F_{\alpha \beta }(x)+\frac
13\varepsilon _\alpha [D_\beta ,F_{\beta \alpha }](x)+{\cal O}(\varepsilon
^2).  \label{a0a1a0}
\end{equation}
With this formula at hand, we have finally
\begin{equation}
G^{{\rm cov}}(x,\varepsilon )=\frac 1{16\pi ^2}\left( \frac 1{|\varepsilon
|^2}+{\rm \ln }m^2|\varepsilon |^2(m^2-\frac{{\rm i}}2\sigma _{\alpha \beta
}F_{\alpha \beta }(x))\right) +{\cal O}(1).
\end{equation}

Before we proceed to the analogous expression for the fermion propagator,
let us now make a useful observation. For any (sufficiently smooth) section $%
\phi (x),$ the operation
\begin{equation}
\phi (x)\rightarrow \Omega (x,x-\varepsilon )\phi (x-\varepsilon )
\end{equation}
can be expressed in the form
\begin{equation}
\Omega (x,x-\varepsilon )\phi (x-\varepsilon )={\rm e}^{-\varepsilon \cdot
\stackrel{\rightarrow }{D_x}}\phi (x).
\end{equation}
Indeed, the functions $\phi _1(x,t)=\Omega (x,x-t\varepsilon )\phi
(x-t\varepsilon )$ and $\phi _2(x,t)={\rm e}^{-t\varepsilon \cdot \stackrel{%
\rightarrow }{D_x}}\phi (x)$ both satisfy the following differential
equation
\begin{equation}
\frac{{\rm d}}{{\rm d}t}\phi _i(x,t)=-\varepsilon \cdot \stackrel{%
\rightarrow }{D_x}\phi _i(x,t)
\end{equation}
with the initial condition
\begin{equation}
\phi _i(x,0)=\phi (x).
\end{equation}
This statement is clear for $i=2$, let us prove it for $i=1$. We have

\begin{eqnarray}
\frac{{\rm d}}{{\rm d}t}\phi _1(x,t) &=&(-\varepsilon \cdot \partial
_y\Omega (x,y))\mid _{y=x-t\varepsilon }\phi (x-t\varepsilon )-\Omega
(x,x-t\varepsilon )\varepsilon \cdot \partial _x\phi (x-t\varepsilon )
\nonumber \\
&=&-\varepsilon \cdot \partial _x(\Omega (x,x-t\varepsilon )\phi
(x-t\varepsilon ))+(\varepsilon \cdot \partial _x\Omega (x,y))\mid
_{y=x-t\varepsilon }\phi (x-t\varepsilon )  \nonumber \\
&=&-\varepsilon \cdot \partial _x(\Omega (x,x-t\varepsilon )\phi
(x-t\varepsilon ))-\varepsilon \cdot A(x)\Omega (x,x-t\varepsilon )\phi
(x-t\varepsilon )  \nonumber \\
&=&-\varepsilon \cdot \stackrel{\rightarrow }{D_x}\phi _1(x,t),
\end{eqnarray}
where we used the relation $(x-y)\cdot D_x\Omega (x,y)=0$ in the third line.

Using now this result and the formula for the fermion propagator

\begin{equation}
S(x,y;m)=-{\rm i}mG(x,y;m^2)+\gamma \cdot \stackrel{\rightarrow }{D}%
_xG(x,y;m^2)
\end{equation}
we have

\[
S^{{\rm cov}}(x,\varepsilon )=-{\rm i}mG^{{\rm cov}}(x,\varepsilon )+{\rm e}%
^{-\varepsilon \cdot \stackrel{\rightarrow }{D_x}}\gamma \cdot \stackrel{%
\rightarrow }{D}_xG(x,y;m^2){\rm e}^{\varepsilon \cdot \stackrel{\leftarrow
}{D}_y}|_{x=y},
\]
where
\begin{eqnarray}
D &=&\partial +A, \\
\stackrel{\leftarrow }{D} &=&\overleftarrow{\partial }-A.
\end{eqnarray}
It is not difficult to show (the details are postponed to Appendix A), that
\begin{eqnarray}
{\rm e}^{-\varepsilon \cdot \stackrel{\rightarrow }{D_x}}\stackrel{%
\rightarrow }{D}_{\mu ,x}G(x,y;m^2){\rm e}^{\varepsilon \cdot \stackrel{%
\leftarrow }{D}_y}|_{x=y} &=&\frac 12([D_\mu ,G^{{\rm cov}}(x,\varepsilon
)]-\partial ^\varepsilon G^{{\rm cov}}(x,\varepsilon )  \nonumber \\
&&\ \ +G_\mu (x,\varepsilon )G^{{\rm cov}}(x,\varepsilon )-G^{{\rm cov}%
}(x,\varepsilon )G_\mu (x,-\varepsilon )  \nonumber \\
&&\ \ +H_\mu (x,\varepsilon )G^{{\rm cov}}(x,\varepsilon )+G^{{\rm cov}%
}(x,\varepsilon )H_\mu (x,-\varepsilon )),  \label{G-H formula}
\end{eqnarray}
where the functions $G_\mu (x,\varepsilon )$ and $H_\mu (x,\varepsilon )$
have the following expansion (the complete formulae are given in Appendix A)
\begin{eqnarray}
G_\mu (x,\varepsilon ) &=&\varepsilon _\nu F_{\mu \nu }(x)-\frac
12[\varepsilon \cdot D,\varepsilon _\nu F_{\mu \nu }(x)]+{\cal O}%
(\varepsilon ^3), \\
H_\mu (x,\varepsilon ) &=&-\frac 12\varepsilon _\nu F_{\nu \mu }(x)+\frac
13[\varepsilon \cdot D,\varepsilon _\nu F_{\nu \mu }(x)]+{\cal O}%
(\varepsilon ^3).
\end{eqnarray}
As a result we have
\begin{eqnarray}
S^{{\rm cov}}(x,\varepsilon ) &=&-{\rm i}mG^{{\rm cov}}(x,\varepsilon
)+\frac 12\gamma _\mu ([D_\mu ,G^{{\rm cov}}(x,\varepsilon )]-\partial _\mu
^\varepsilon G^{{\rm cov}}(x,\varepsilon )  \nonumber \\
&&\ \ \ +G_\mu (x,\varepsilon )G^{{\rm cov}}(x,\varepsilon )-G^{{\rm cov}%
}(x,\varepsilon )G_\mu (x,-\varepsilon )  \nonumber \\
&&\ \ \ +H_\mu (x,\varepsilon )G^{{\rm cov}}(x,\varepsilon )+G^{{\rm cov}%
}(x,\varepsilon )H_\mu (x,-\varepsilon )).
\end{eqnarray}
and, putting the formulas together,
\begin{eqnarray}
S^{{\rm cov}}(x,\varepsilon ) &=&\frac 1{16\pi ^2}\left(\frac{\gamma \cdot
\varepsilon }{|\varepsilon |^4}-\frac{{\rm i}m}{|\varepsilon |^2}-\frac{%
m^2\gamma \cdot \varepsilon }{|\varepsilon |^2}-\frac{\varepsilon _\mu
F_{\mu \nu }^{*}(x)\gamma _\nu \gamma _5}{|\varepsilon |^2}
\right.
\nonumber \\
&&\ \ \ +{\rm \ln }m^2|\varepsilon |^2(-{\rm i}m^3-\frac 12m\sigma _{\alpha
\beta }F_{\alpha \beta }(x)-\frac 13[D_\mu ,F_{\mu \nu }]\gamma _\nu +{\cal O%
}(\varepsilon ))
\nonumber \\
&&
\left.
\ \ \ +\frac 13\frac{\gamma \cdot \varepsilon [D_\mu ,F_{\mu \nu
}]\varepsilon _\nu }{|\varepsilon |^2}-\frac 13\frac{\gamma _\mu
[\varepsilon \cdot D,F_{\mu \nu }]\varepsilon _\nu }{|\varepsilon |^2}+{\cal %
O}_{{\rm reg}}(1)\right).  \label{covariant singularity}
\end{eqnarray}
Here we keep for the sake of further convenience also the terms, which are
formally ${\cal O}(1),$ but the $\varepsilon \rightarrow 0$ limit of which
does not exist, but rather depends on the direction in which $\varepsilon $
approaches zero. As we shall see in the following, this terms are
responsible for the anomalous divergence of the energy-momentum tensor.


\section{The anomalies via point-splitting \label{anomalies}}


The quantum anomalies generally mean a violation of the classical symmetry
due to the quantum corrections, which cannot be avoided by suitable
renormalization of the quantities involved. Historically, the first example
of such a phenomenon was the famous Adler-Bell-Jackiw chiral anomaly \cite
{ABJhist}. It is connected with the so-called chiral $U(1)$ transformation
\begin{eqnarray}
\delta \psi &=&{\rm i}\alpha \gamma _5\psi  \nonumber \\
\delta \psi ^{+} &=&{\rm i}\psi ^{+}\alpha \gamma _5,
\end{eqnarray}
where $\alpha $ is an infinitesimal real parameter. The corresponding
Noether current
\begin{equation}
j_{\mu 5}(x)=\psi ^{+}\gamma _\mu \gamma _5\psi
\end{equation}
obeys classically the following equation
\begin{equation}
\partial \cdot j_5(x)=2{\rm i}m\psi ^{+}\gamma _5\psi
\end{equation}
and it is conserved in the chiral limit $m\rightarrow 0.$ In the quantum
case, however, the anomalous term $\frac 1{8\pi ^2}F_{\mu \nu }^{*}F_{\mu
\nu }$, (where $F_{\mu \nu }^{*}$ is the dual tensor to $F_{\mu \nu }$)
appears on the right hand side of the previous equation, spoiling the chiral
invariance of the theory. Though this term could be removed by a suitable
counterterm added to the chiral current $j_{\mu 5}(x),$ this counterterm
spoils gauge invariance and is therefore not admissible in any reasonable
gauge theory. This situation is typical; the anomalous symmetries appear in
pairs and saving one of them necessarily brings about spoiling of the other.
Further example of the anomalous pair is the conflict between the scale
and translational symmetries leading to the so called trace anomaly\footnote{%
It results in the anomalous trace of the energy-momentum tensor.}\cite{trace}.
In the gravitational background there is
also the so-called Lorentz anomaly, the consequence of which is the
anomalous antisymmetric part of the energy-momentum tensor. For a
comprehensive discussion of this and related topics see the books \cite
{Bertlmann}, where also an extensive list of references can be found.

In this section we give rather nonstandard derivation of the axial and trace
anomalies via point-splitting regularization. We will show, that the
anomalies in this formalism can be understood as the result of the
non-invariance of the classical action with respect to the regularized
nonlocal ``point-split'' transformations, which replace the naive form of
the chiral rotation, scale transformation and (covariant) translation. The
remnant of this noninvariance survives in the quantum case the procedure of
removing the cut-off. This gives a partial explanation of why the
point-splitting regularization produces anomaly in the divergence rather
than in the trace of the energy-momentum tensor. The idea of the deformed
transformations was used in similar context of the Fujikawa-like
regularization procedures in the paper \cite{Joglekar}, where the anomalous
pair of the chiral and abelian gauge symmetry was investigated.

The main object of our interest is the Euclidean generating functional of
the correlators of the gauge fermionic currents given formally as the
fermionic functional integral
\begin{equation}
Z_E[A]=\int {\cal D}\psi ^{+}{\cal D}\psi \exp (-S_E),.
\end{equation}
where
\begin{equation}
S_E=\int {\rm d}^4x\psi ^{+}(\frac 12\gamma \cdot \stackrel{%
\longleftrightarrow }{D}+{\rm i}m)\psi
\end{equation}
is the Euclidean action. We shall tacitly assume some intermediate
regularization (e.g. Pauli-Villars) with the cut-off parameter $\Lambda $,
which makes the functional integral well defined and which justifies the
formal operations we are going to perform. This intermediate regularization
will be removed at the end of the calculation. The only requirements on this
regularization are that it does not deform the four-dimensional algebra of
the $\gamma -$matrices (in order to preserve the four-dimensional $\gamma _5$
) and also the property which we call ``naturalness'' - i.e. that it
reproduces the true value of the convergent integrals after the cut-off is
removed. The Ward-Takahashi identity, corresponding to the general
infinitesimal change of the integration variables
\begin{eqnarray}
\psi &\rightarrow &\psi +\delta \psi  \nonumber \\
\psi ^{+} &\rightarrow &\psi ^{+}+\delta \psi ^{+}  \label{transformation}
\end{eqnarray}
has the form
\begin{equation}
\int {\cal D}\psi ^{+}{\cal D}\psi \left( \delta S_E+\int {\rm d}^4x\left(
\frac{\delta \delta \psi (x)}{\delta \psi (x)}+\frac{\delta \delta \psi
^{+}(x)}{\delta \psi ^{+}(x)}\right) \right) \exp (-S_E)=0,
\label{W-T global}
\end{equation}
where $\delta S_E=\int {\rm d}^4x\delta {\cal L}_E(x)$ is the variation of
the action with respect to the transformation (\ref{transformation}). Here
we pick up the formal remnant of the determinant of the change of the
integration variables, which we assume to be consistently regularized by the
above mentioned intermediate regularization\footnote{%
E.g. for the Pauli-Villars regularization this formally divergent expression
is cancelled by the analogous term coming from the transformation of the
regulator fields with opposite statistics. Within the framework of the
dimensional regularization it vanishes by definition.}. Within the naive
form of the Ward-Takahashi identity this term is omitted, on the other hand
it can be the source of the quantum anomaly provided it survives after the
cut-off is removed\footnote{%
However, for the Pauli-Villars regularization, the source of the axial
anomaly is the additional contribution of the regulator fields to $\delta
S_E $. Within the consistent dimensional regularization in this case also
the additional terms appear in $\delta S_E$ due to the commutation of $%
\gamma _5$ with the $\gamma _\mu $ with $\mu >4$.}. Note that the local form
of the Ward-Takahashi identity can be obtained by means of the standard
technique of localizing the transformation (\ref{transformation}) and can be
written in the form
\begin{equation}
\int {\cal D}\psi ^{+}{\cal D}\psi \left( \int {\rm d}^4x\theta (x)\left(
\delta {\cal L}_E(x)-\partial \cdot j(x)+\left( \frac{\delta \delta \psi (x)%
}{\delta \psi (x)}+\frac{\delta \delta \psi ^{+}(x)}{\delta \psi ^{+}(x)}%
\right) \right) \right) \exp (-S_E)=0,  \label{W-T local}
\end{equation}
where $\theta (x)$ is an infinitesimal parameter of the localized
transformation
\begin{eqnarray}
\psi &\rightarrow &\psi +\theta \delta \psi  \nonumber \\
\psi ^{+} &\rightarrow &\psi ^{+}+\theta \delta \psi ^{+}
\label{transformation local}
\end{eqnarray}
and $j_\mu (x)$ is the Noether current. Another subject of our investigation
will be therefore the generating functional of the connected correlators of
the gauge currents with one insertion of the divergence of the Noether
current $j_\mu (x)$ or one insertion of the $\delta {\cal L}_E(x)$
corresponding to the transformation (\ref{transformation}), assuming these
operators to be regularized by point-splitting, i.e.
\begin{eqnarray}
\partial \cdot j(x)^{{\rm reg}}\cdot Z_E[A]_c &=&\langle \partial \cdot
j(x)^{{\rm reg}}\rangle _c=\int {\cal D}\psi ^{+}{\cal D}\psi \,j_\mu (x)^{%
{\rm reg}}\exp (-S_E)|_c  \nonumber \\
&=&Z_E[A]^{-1}\int {\cal D}\psi ^{+}{\cal D}\psi \,j_\mu (x)^{{\rm reg}}\exp
(-S_E)
\end{eqnarray}
and
\begin{eqnarray}
\delta {\cal L}_E(x)^{{\rm reg}}\cdot Z_E[A]_c &=&\langle \delta {\cal L}%
_E(x)^{{\rm reg}}\rangle _c=\int {\cal D}\psi ^{+}{\cal D}\psi \,\delta
{\cal L}_E(x)^{{\rm reg}}\exp (-S_E)|_c  \nonumber \\
&=&Z_E[A]^{-1}\int {\cal D}\psi ^{+}{\cal D}\psi \,\delta {\cal L}_E(x)^{%
{\rm reg}}\exp (-S_E).
\end{eqnarray}
Here we need not assume any other intermediate regularization, since the
loops with one insertion of the regularized operator are already finite and
the loops without the operator insertion are formally cancelled.
Alternatively we can think about this point-split functionals as being
produced by removing the intermediate cut-off, which is possible because all
the relevant integrals are finite and thanks to the required naturalness
property of the intermediate regularization are not deformed by this
procedure. This is the point of view we will adopt in the following.

In the point-splitting method we introduce the anomaly as the insertion
defined by the difference
\begin{equation}
{\cal A}(x)\cdot Z_E[A]_c=\lim_{{\rm removed\;point-splitting}}\left(
\partial \cdot j(x)^{{\rm reg}}\cdot Z_E[A]_c-\delta {\cal L}_E(x)^{{\rm reg}%
}\cdot Z_E[A]_c\right) ,  \label{general anomaly}
\end{equation}
i.e. as the anomalous divergence of the Noether current in the limit of the
point-splitting cut-off removed. Our strategy for calculation of the
anomalies will be as follows. We first deform the localized transformation
prescription (\ref{transformation local}) in such a way, that the
corresponding Noether current is identical with the point-split Noether
current of the original transformation. As a result, the local
Ward-Takahashi identity (\ref{W-T local}) for the deformed transformation
will produce the point-split version of the original identity, i.e. with the
relevant composite operators regularized by the point-splitting. As we shall
see, the deformation of the transformation brings about the presence of
additional terms in $\delta S_E$ which prove to be the source of the
anomaly. On the other hand, the determinant of the deformed transformation
will be identically equal to one; this property also survives the process of
removing of the intermediate cut-off. In the next sections we give explicit
examples of this general strategy.


\section{The chiral anomaly}


Let us give the first illustration of the above described strategy. For the
nonabelian chiral transformation
\begin{eqnarray}
\delta \psi &=&\alpha \gamma _5\psi  \nonumber \\
\delta \psi ^{+} &=&\psi ^{+}\alpha \gamma _5  \label{chiral rotation}
\end{eqnarray}
where $\alpha =\alpha _aT^a$ is an antihermitian infinitesimal matrix from
the Lie algebra of the gauge group, the Noether currents are identical with
the axial currents given by
\begin{equation}
j_{5\mu }^a=\psi ^{+}T^a\gamma _\mu \gamma _5\psi
\end{equation}
and on the classical level we have
\begin{equation}
D_\mu j_{5\mu }^a=\delta {\cal L}_E(x)^a=2{\rm i}m\psi ^{+}T^a\gamma _5\psi ,
\end{equation}
where $\delta {\cal L}_E(x)^a$ is the variation of the Lagrangian under the
chiral rotations. Our choice for the point-split version of these currents is%
\footnote{%
There are also other possibilities used in the literature, e.g.
\[
j_{5\mu }^{a,{\rm reg}}=\psi ^{+}(x+\varepsilon )\{\Omega (x+\varepsilon
,x-\varepsilon ),T^a\}\gamma _\mu \gamma _5\psi (x-\varepsilon ),
\]
which has also covariant transformation properties. Our choice was motivated
mainly by the relative simplicity of the further calculation.}
\begin{equation}
j_{5\mu }^{a,{\rm reg}}=\psi ^{+}(x+\varepsilon )\Omega (x+\varepsilon
,x)T^a\gamma _\mu \gamma _5\Omega (x,x-\varepsilon )\psi (x-\varepsilon ),
\end{equation}
where $\Omega (x,y)$ is the parallel transporter along the straight line
connecting the points $x$ and $y$. Note, that this definition gives gauge
covariant expression; the regularized current multiplet transforms according
to the adjoint representation of the gauge group. For the further
convenience, let us introduce also the point-split fields (cf. also Section
\ref{asymptotic})
\begin{eqnarray}
\psi _\varepsilon (x) &=&{\rm e}^{-\varepsilon \cdot D}\psi (x)=\Omega
(x,x-\varepsilon )\psi (x-\varepsilon ),  \nonumber \\
\psi _\varepsilon ^{+}(x) &=&\psi ^{+}(x){\rm e}^{\varepsilon \cdot
\stackrel{\leftarrow }{D}}=\psi ^{+}(x+\varepsilon )\Omega (x+\varepsilon
,x),
\end{eqnarray}
in terms of which
\begin{equation}
j_{5\mu }^{a,{\rm reg}}=\psi _\varepsilon ^{+}T^a\gamma _\mu \gamma _5\psi
_\varepsilon .
\end{equation}
Let us note, that the fields $\psi _\varepsilon (x)$ and $\psi _\varepsilon
^{+}(x)$ have the same transformation properties with respect to the gauge
transformations as the original fields $\psi (x)$ and $\psi ^{+}(x)$. As a
result, the regularized current $j_{5\mu }^{a,{\rm reg}}$ is gauge
covariant. Note also, that the covariant propagator introduced in the
previous section can be expressed in terms of the correlator of the
point-split fields
\begin{equation}
\langle \psi _\varepsilon (x)\psi _\varepsilon ^{+}(x)\rangle
_c=Z_E[A]^{-1}\int {\cal D}\psi ^{+}{\cal D}\psi \,\psi _\varepsilon (x)\psi
_\varepsilon ^{+}(x)\exp (-S_E)=S^{{\rm cov}}(x,\varepsilon ).
\label{correlator}
\end{equation}
The variation of the Lagrangian under the chiral rotations, which represent
the naive divergence of the axial currents
\begin{equation}
\delta {\cal L}_E(x)^a=2{\rm i}m\psi ^{+}T^a\gamma _5\psi
\end{equation}
can be regulated analogously as
\[
\delta {\cal L}_E(x)^{a,{\rm reg}}=2{\rm i}m\psi ^{+}(x+\varepsilon )\Omega
(x+\varepsilon ,x)T^a\gamma _5\Omega (x,x-\varepsilon )\psi (x-\varepsilon
)=2{\rm i}m\psi _\varepsilon ^{+}T^a\gamma _5\psi _\varepsilon .
\]

Let us now introduce the deformed local chiral rotation with infinitesimal
parameter $\alpha (x)=\alpha _a(x)T^a$as follows
\begin{eqnarray}
\delta \psi &=&{\rm e}^{-\varepsilon \cdot D}\alpha \gamma _5{\rm e}%
^{-\varepsilon \cdot D}\psi ,  \nonumber \\
\delta \psi ^{+} &=&\psi ^{+}{\rm e}^{\varepsilon \cdot \stackrel{\leftarrow
}{D}}\alpha \gamma _5{\rm e}^{\varepsilon \cdot \stackrel{\leftarrow }{D}}.
\label{deformed chiral rotation}
\end{eqnarray}
Because of the shift of the argument in the expression for $\delta \psi $
and $\delta \psi ^{+}$, we have

\begin{equation}
\int {\rm d}^4x\left( \frac{\delta \delta \psi (x)}{\delta \psi (x)}+\frac{%
\delta \delta \psi ^{+}(x)}{\delta \psi ^{+}(x)}\right) =0
\end{equation}
independently of the intermediate regularization. The change of the
Euclidean action under the deformed local chiral rotation (\ref
{deformed
chiral rotation}) is

\begin{eqnarray}
\delta S_E &=&\int {\rm d}^4x\left( \psi _\varepsilon ^{+}[\gamma \cdot
D,\alpha ]\gamma _5\psi _\varepsilon +2{\rm i}m\psi _\varepsilon ^{+}\alpha
\gamma _5\psi _\varepsilon \right) \\
&&\ +\int {\rm d}^4x\left( \psi _\varepsilon ^{+}(\alpha \gamma _5[{\rm e}%
^{-\varepsilon \cdot \stackrel{\rightarrow }{D}},\gamma \cdot \stackrel{%
\rightarrow }{D}]{\rm e}^{\varepsilon \cdot \stackrel{\rightarrow }{D}}-{\rm %
e}^{-\varepsilon \cdot \stackrel{\leftarrow }{D}}[\gamma \cdot \stackrel{%
\leftarrow }{D},{\rm e}^{\varepsilon \cdot \stackrel{\leftarrow }{D}}]\gamma
_5\alpha )\psi _\varepsilon \right) .  \nonumber
\end{eqnarray}
Inserting this to the Ward-Takahashi identity (\ref{W-T global}) and
removing the intermediate cut-off we get for the covariant divergence of the
point-split axial current

\begin{equation}
D_\mu \langle j_{5\mu }^{a,{\rm reg}}\rangle _c=2{\rm i}m\langle \psi
_\varepsilon ^{+}T^a\gamma _5\psi _\varepsilon \rangle _c+{\cal A}_5^a,
\end{equation}
where the point-split axial anomaly is given by the formula (cf. (\ref
{general anomaly}) and (\ref{correlator}))

\begin{eqnarray}
{\cal A}_5^a &=&\langle \psi _\varepsilon ^{+}(T^a\gamma _5[{\rm e}%
^{-\varepsilon \cdot \stackrel{\rightarrow }{D}},\gamma \cdot \stackrel{%
\rightarrow }{D}]{\rm e}^{\varepsilon \cdot \stackrel{\rightarrow }{D}}-{\rm %
e}^{-\varepsilon \cdot \stackrel{\leftarrow }{D}}[\gamma \cdot \stackrel{%
\leftarrow }{D},{\rm e}^{\varepsilon \cdot \stackrel{\leftarrow }{D}%
}]T^a\gamma _5)\psi _\varepsilon \rangle _c  \nonumber \\
&=&-{\rm Tr}S^{{\rm cov}}\gamma _5(T^a[{\rm e}^{-\varepsilon \cdot \stackrel{%
\rightarrow }{D}},\gamma \cdot \stackrel{\rightarrow }{D}]{\rm e}%
^{\varepsilon \cdot \stackrel{\rightarrow }{D}}+{\rm e}^{-\varepsilon \cdot
\stackrel{\leftarrow }{D}}[\gamma \cdot \stackrel{\leftarrow }{D},{\rm e}%
^{\varepsilon \cdot \stackrel{\leftarrow }{D}}]T^a);
\end{eqnarray}
here the trace is taken over both the Dirac and color indices. Remembering
the formulae
\begin{eqnarray}
\lbrack {\rm e}^{-\varepsilon \cdot D},D_\mu ]{\rm e}^{\varepsilon \cdot
D}=G_\mu (x,\varepsilon )
\end{eqnarray}
and

\begin{eqnarray}
{\rm e}^{-\varepsilon \cdot \stackrel{\leftarrow }{D}}[\stackrel{\leftarrow
}{D_\mu },{\rm e}^{\varepsilon \cdot \stackrel{\leftarrow }{D}}]=-G_\mu
(x,-\varepsilon )
\end{eqnarray}
we get for the anomaly
\begin{equation}
{\cal A}_5^a=-{\rm Tr}S^{{\rm cov}}\gamma _5([T^a,\gamma \cdot
G^{+}]+\{T^a,\gamma \cdot G^{-}\})  \label{A anomaly}
\end{equation}
where
\begin{equation}
G_\mu ^{\pm }(x,\varepsilon )=\frac 12(G_\mu (x,\varepsilon )\pm G_\mu
(x,-\varepsilon )),  \label{Gpm}
\end{equation}
i.e. (cf. also Appendix A)
\begin{eqnarray}
G_\mu ^{+}(x,\varepsilon ) &=&-\frac 12[\varepsilon \cdot D,F_{\mu \nu
}]\varepsilon _\nu +{\cal O}(|\varepsilon |^4),  \label{G+} \\
G_\mu ^{-}(x,\varepsilon ) &=&\varepsilon _\nu F_{\mu \nu }(x)+{\cal O}%
(|\varepsilon |^3).  \label{G-}
\end{eqnarray}
From the expression for the short distance behavior of the covariant
propagator we see that only the second term on the right hand side of (\ref
{A anomaly}) contributes in the limit $\varepsilon \rightarrow 0$
\begin{eqnarray}
{\cal A}_5^a=-{\rm Tr}\left( -\frac 1{16\pi ^2}
\left(\frac{\varepsilon _\mu }{|\varepsilon |^2}
F_{\mu \nu }^{*}\gamma _\nu \gamma _\alpha
+{\cal O}(1)\right)
(\{T^a,F_{\alpha \sigma }\varepsilon _\sigma \}+{\cal O}(\varepsilon
^2))\right) .
\end{eqnarray}
Taking the average over the directions of the four-vector $\varepsilon _\mu $
and performing the trace of the Dirac matrices we get finally the covariant
form of the axial anomaly

\begin{equation}
{\cal A}_5^a=\frac 1{8\pi ^2}{\rm Tr}_CT^aF_{\mu \nu }^{*}F_{\mu \nu }+{\cal %
O}(\varepsilon ),  \label{axial anomaly}
\end{equation}
where ${\rm Tr}_C$ means a trace over the color indices only.


\section{The vector anomaly}


As another simple application of the above described method let us consider
the nonabelian gauge transformation
\begin{eqnarray}
\delta \psi &=&\alpha \psi  \nonumber \\
\delta \psi ^{+} &=&-\psi ^{+}\alpha  \label{gauge rotation}
\end{eqnarray}
where again $\alpha =\alpha _aT^a$. The corresponding Noether currents
\begin{equation}
j_\mu ^a=\psi ^{+}T^a\gamma _\mu \psi
\end{equation}
are covariantly conserved at the classical level, i.e.
\begin{equation}
D_\mu j_\mu ^a=0.
\end{equation}
Let us choose the following point-split version of these vector currents%
\footnote{%
Here again the regularized current $j_\mu ^{a,{\rm reg}}$ is gauge covariant
because of the transformation properties of the point-split fields $\psi
_\varepsilon $ and $\psi _\varepsilon ^{+}$.}
\begin{equation}
j_\mu ^{a,{\rm reg}}=\psi _\varepsilon ^{+}T^a\gamma _\mu \psi _\varepsilon .
\end{equation}
Because the gauge covariance is manifest within our point-splitting
regularization scheme, we do not expect any anomalous divergence. This can
be easily proved using the technique described above. Let us consider the
point-split transformation with infinitesimal parameter $\alpha (x)=\alpha
_a(x)T^a$%
\begin{eqnarray}
\delta \psi &=&{\rm e}^{-\varepsilon \cdot D}\alpha {\rm e}^{-\varepsilon
\cdot D}\psi ,  \nonumber \\
\delta \psi ^{+} &=&-\psi ^{+}{\rm e}^{\varepsilon \cdot \stackrel{%
\leftarrow }{D}}\alpha {\rm e}^{\varepsilon \cdot \stackrel{\leftarrow }{D}}.
\label{deformed gauge rotation}
\end{eqnarray}
The variation of the action is then
\begin{eqnarray}
\delta S_E=\int {\rm d}^4x\left( \psi _\varepsilon ^{+}[\gamma \cdot
D,\alpha ]\psi _\varepsilon +\psi _\varepsilon ^{+}(\alpha [{\rm e}%
^{-\varepsilon \cdot \stackrel{\rightarrow }{D}},\gamma \cdot \stackrel{%
\rightarrow }{D}]{\rm e}^{\varepsilon \cdot \stackrel{\rightarrow }{D}}+{\rm %
e}^{-\varepsilon \cdot \stackrel{\leftarrow }{D}}[\gamma \cdot \stackrel{%
\leftarrow }{D},{\rm e}^{\varepsilon \cdot \stackrel{\leftarrow }{D}}]\alpha
)\psi _\varepsilon \right) .
\end{eqnarray}
We have therefore the following point-split Ward-Takahashi identity
\begin{equation}
D_\mu \langle j_\mu ^{a,{\rm reg}}\rangle ={\cal A}^a
\end{equation}
where the possible vector anomaly is
\begin{equation}
{\cal A}^a=\langle \psi _\varepsilon ^{+}(T^a[{\rm e}^{-\varepsilon \cdot
\stackrel{\rightarrow }{D}},\gamma \cdot \stackrel{\rightarrow }{D}]{\rm e}%
^{\varepsilon \cdot \stackrel{\rightarrow }{D}}+{\rm e}^{-\varepsilon \cdot
\stackrel{\leftarrow }{D}}[\gamma \cdot \stackrel{\leftarrow }{D},{\rm e}%
^{\varepsilon \cdot \stackrel{\leftarrow }{D}}]T^a)\psi _\varepsilon \rangle
.
\end{equation}
This can be rewritten in the form
\begin{eqnarray}
{\cal A}^a=-{\rm Tr}S^{{\rm cov}}(T^aG(x,\varepsilon )\cdot \gamma -\gamma
\cdot G(x,-\varepsilon )T^a)
\end{eqnarray}
where $G_\mu (x,\varepsilon )$ was introduced in the previous section. Note,
that $\varepsilon \cdot G(x,\varepsilon )=0$ and $G_\mu (x,\varepsilon )=%
{\cal O}(\varepsilon )$. Using the formula for the covariant propagator,
which implies
\begin{eqnarray}
{\rm Tr}_DS^{{\rm cov}}\gamma _\mu &=&\frac 1{4\pi ^2}\left\{-\frac{\varepsilon
_\mu }{|\varepsilon |^4}+m^2\frac{\varepsilon _\mu }{|\varepsilon |^2}+\frac
13{\rm \ln }m^2|\varepsilon |^2[D_\sigma ,F_{\sigma \mu }]
\right.
\nonumber \\
&&
\left.
\ \ -\frac 13\frac{\varepsilon _\mu \varepsilon _\sigma }{|\varepsilon |^2%
}[D_\rho ,F_{\rho \sigma }]+\frac 13\frac{\varepsilon _\rho \varepsilon
_\sigma }{|\varepsilon |^2}[D_\rho ,F_{\mu \sigma }]+{\cal O}_{{\rm reg}%
}(1)\right\},  \label{TrSgamma}
\end{eqnarray}
we get
\begin{equation}
{\cal A}^a=-\frac 1{4\pi ^2}\left(-\frac{\varepsilon _\mu }{|\varepsilon |^4}+m^2%
\frac{\varepsilon _\mu }{|\varepsilon |^2}+
{\cal O}(\ln |\varepsilon|^2)\right)%
{\rm Tr}_C(T^aG_\mu (x,\varepsilon )-G_\mu (x,-\varepsilon )T^a)={\cal O}%
(\varepsilon \ln |\varepsilon |^2)
\end{equation}
and therefore the vector current is anomaly free in the limit of the removed
point-splitting.


\section{The trace anomaly}


Another example of the strategy of deformed transformation is the trace
anomaly of the canonical gauge invariant energy-momentum tensor\footnote{%
This tensor is not symmetric; the construction of the symmetric
energy-momentum tensor will be discussed in the next section.}
\begin{equation}
\theta _{\mu \nu }=\frac 12\psi ^{+}\gamma _\mu \stackrel{\leftrightarrow }{D%
}_\nu \psi .  \label{E-M}
\end{equation}
Let us consider the scale transformation
\begin{eqnarray}
\delta \psi &=&\lambda \psi ,  \nonumber \\
\delta \psi ^{+} &=&\psi ^{+}\lambda ,  \label{scale transformation}
\end{eqnarray}
where $\lambda $ is a real infinitesimal parameter. The corresponding
Noether current is identically zero and the change of the action under
localized scale transformation with parameter $\lambda (x)$ is
\begin{equation}
\delta S_E=\int {\rm d}^4x\,2\lambda (x)\left( \frac 12\psi ^{+}\gamma \cdot
\stackrel{\leftrightarrow }{D}\psi +{\rm i}m\psi ^{+}\psi \right) =\int {\rm %
d}^4x\,2\lambda (x)\left( \theta _{\mu \mu }+{\rm i}m\psi ^{+}\psi \right) .
\label{scale change}
\end{equation}
On the classical level we have therefore
\begin{equation}
\theta _{\mu \mu }=-{\rm i}m\psi ^{+}\psi .
\end{equation}
The regularized form of the relevant composite operators is\footnote{%
Other forms of the point-split tensor $\theta _{\mu \nu }^{{\rm reg}}$ are
used in the literature as well. Our choice is again motivated by simplicity.}
\begin{equation}
\theta _{\mu \nu }^{{\rm reg}}=\frac 12\psi _\varepsilon ^{+}\gamma _\mu
\stackrel{\leftrightarrow }{D}_\nu \psi _\varepsilon  \label{E-M reg}
\end{equation}
and
\begin{equation}
\psi ^{+}\psi ^{{\rm reg}}=\psi _\varepsilon ^{+}\psi _\varepsilon ;
\label{S reg}
\end{equation}
these regularized expressions are gauge invariant. The point-split version
of the transformation (\ref{scale transformation}) is then

\begin{eqnarray}
\delta \psi &=&{\rm e}^{-\varepsilon \cdot D}\lambda {\rm e}^{-\varepsilon
\cdot D}\psi ,  \nonumber \\
\delta \psi ^{+} &=&\psi ^{+}{\rm e}^{\varepsilon \cdot \stackrel{\leftarrow
}{D}}\lambda {\rm e}^{\varepsilon \cdot \stackrel{\leftarrow }{D}}.
\label{deformed scale transformation}
\end{eqnarray}
The Jacobian of this deformed scaling is identically equal to one for the
same reason as before. The variation of the action reproduces the
point-split form of (\ref{scale change}) and acquires an additional term,
which is the potential source of the trace anomaly

\begin{eqnarray}
\delta S_E &=&\int {\rm d}^4x\,2\lambda \left( \frac 12\psi _\varepsilon
^{+}\gamma \cdot \stackrel{\leftrightarrow }{D}\psi _\varepsilon +{\rm i}%
m\psi _\varepsilon ^{+}\psi _\varepsilon \right)  \nonumber \\
&&+\int {\rm d}^4x\lambda \left( \psi _\varepsilon ^{+}([{\rm e}%
^{-\varepsilon \cdot \stackrel{\rightarrow }{D}},\gamma \cdot \stackrel{%
\rightarrow }{D}]{\rm e}^{\varepsilon \cdot \stackrel{\rightarrow }{D}}-{\rm %
e}^{-\varepsilon \cdot \stackrel{\leftarrow }{D}}[\gamma \cdot \stackrel{%
\leftarrow }{D},{\rm e}^{\varepsilon \cdot \stackrel{\leftarrow }{D}}])\psi
_\varepsilon \right) ,
\end{eqnarray}
i.e. we have

\begin{equation}
\langle \theta _{\mu \mu }^{{\rm reg}}\rangle _c=-{\rm i}\langle m\psi
_\varepsilon ^{+}\psi _\varepsilon \rangle _c+{\cal A}^{{\rm trace}}
\end{equation}
where
\begin{eqnarray}
{\cal A}^{{\rm trace}} &=&-\frac 12\langle \psi _\varepsilon ^{+}([{\rm e}%
^{-\varepsilon \cdot \stackrel{\rightarrow }{D}},\gamma \cdot \stackrel{%
\rightarrow }{D}]{\rm e}^{\varepsilon \cdot \stackrel{\rightarrow }{D}}-{\rm %
e}^{-\varepsilon \cdot \stackrel{\leftarrow }{D}}[\gamma \cdot \stackrel{%
\leftarrow }{D},{\rm e}^{\varepsilon \cdot \stackrel{\leftarrow }{D}}])\psi
_\varepsilon \rangle _c  \nonumber \\
\ &=&{\rm Tr}S^{{\rm cov}}\gamma \cdot G^{+},
\end{eqnarray}
where $G_\mu ^{+}$ was introduced via the formula (\ref{Gpm}). Using the
formulae (\ref{TrSgamma}) and (\ref{G+}) we get

\begin{equation}
{\cal A}^{{\rm trace}}={\rm Tr}_C\left( \frac 1{16\pi ^2}\left(-4\frac{%
\varepsilon _\mu }{|\varepsilon |^4}+{\cal O}(\frac{\varepsilon _\mu }{%
|\varepsilon |^2})\right) (-\frac 12[\varepsilon \cdot D,F_{\mu \nu }]\varepsilon
_\nu +{\cal O}(|\varepsilon |^4))\right) ,
\end{equation}
and, after taking the average over the direction of the four-vector $%
\varepsilon _\mu $,

\begin{equation}
{\cal A}^{{\rm trace}}={\cal O}(\varepsilon ).
\end{equation}
It means that there is no trace anomaly within the point-splitting
regularization.


\section{The translation anomaly}


Because there is no anomaly in the trace of the point-split energy-momentum
tensor, we expect the appearance of an anomaly in its divergence. Assume
therefore the covariant translation of the fields, parametrized by
an infinitesimal four-vector $a$%
\begin{eqnarray}
\delta \psi &=&a\cdot D\psi ,  \nonumber \\
\delta \psi ^{+} &=&\psi ^{+}\stackrel{\leftarrow }{D}\cdot a.
\end{eqnarray}
The canonical Noether current associated with this transformation is
identical with the energy-momentum tensor (\ref{E-M}) introduced above. The
variation of the action with respect to the localized version of this
transformation, which gives the classical divergence of the energy-momentum
tensor, is
\begin{eqnarray}
\delta S_E &=&\int {\rm d}^4x\left[
\begin{array}{c}
-a_\nu \partial _\mu \left( \frac 12\psi ^{+}\gamma _\mu \stackrel{%
\leftrightarrow }{D}_\nu \psi \right) +a\cdot \partial \left( \frac 12\psi
^{+}\gamma \cdot \stackrel{\leftrightarrow }{D}\psi +{\rm i}m\psi ^{+}\psi
\right) +a_\nu \psi ^{+}\gamma _\mu F_{\mu \nu }\psi
\end{array}
\right]  \nonumber \\
&=&\int {\rm d}^4x\left[ -a_\nu \partial _\mu \theta _{\mu \nu }+a\cdot
\partial \left( \theta _{\mu \mu }+{\rm i}m\psi ^{+}\psi \right) +a_\nu \psi
^{+}\gamma _\mu F_{\mu \nu }\psi \right] ,
\end{eqnarray}
on the classical level we have therefore
\begin{equation}
\partial _\mu \theta _{\mu \nu }=\psi ^{+}\gamma _\mu F_{\mu \nu }\psi .
\end{equation}
We again regularize the operators involved according to the (\ref{E-M reg}),
(\ref{S reg}) and to the prescription

\begin{equation}
\psi ^{+}\gamma _\mu F_{\mu \nu }\psi ^{{\rm reg}}=\psi _\varepsilon
^{+}\gamma _\mu F_{\mu \nu }\psi _\varepsilon .
\end{equation}
In order to get the Ward-Takahashi identity for the point-split operators,
we deform the (local) covariant translation as follows
\begin{eqnarray}
\delta \psi &=&{\rm e}^{-\varepsilon \cdot D}a\cdot D{\rm e}^{-\varepsilon
\cdot D}\psi  \nonumber \\
\delta \psi ^{+} &=&\psi ^{+}{\rm e}^{\varepsilon \cdot \stackrel{\leftarrow
}{D}}\stackrel{\leftarrow }{D}\cdot a{\rm e}^{\varepsilon \cdot \stackrel{%
\leftarrow }{D}}  \label{deformed translation}
\end{eqnarray}
We have then for the variation of the action

\begin{eqnarray}
\delta S_E &=&\int {\rm d}^4x\left[
\begin{array}{c}
-a_\nu \partial _\mu \left( \frac 12\psi _\varepsilon ^{+}\gamma _\mu
\stackrel{\leftrightarrow }{D}_\nu \psi _\varepsilon \right) +a\cdot
\partial \left( \frac 12\psi _\varepsilon ^{+}\gamma \cdot \stackrel{%
\leftrightarrow }{D}\psi _\varepsilon +{\rm i}m\psi _\varepsilon ^{+}\psi
_\varepsilon \right) +a_\nu \psi _\varepsilon ^{+}\gamma _\mu F_{\mu \nu
}\psi _\varepsilon
\end{array}
\right]  \nonumber \\
&&+\int {\rm d}^4xa_\nu \left( \psi _\varepsilon ^{+}(\stackrel{\leftarrow }{%
D}_\nu [{\rm e}^{-\varepsilon \cdot \stackrel{\rightarrow }{D}},\gamma \cdot
\stackrel{\rightarrow }{D}]{\rm e}^{\varepsilon \cdot \stackrel{\rightarrow
}{D}}-{\rm e}^{-\varepsilon \cdot \stackrel{\leftarrow }{D}}[\gamma \cdot
\stackrel{\leftarrow }{D},{\rm e}^{\varepsilon \cdot \stackrel{\leftarrow }{D%
}}]\stackrel{\rightarrow }{D}_\nu )\psi _\varepsilon \right)
\end{eqnarray}
and therefore

\begin{eqnarray}
\partial _\mu \langle \theta _{\mu \nu }^{{\rm reg}}\rangle _c=\partial _\nu
\langle \frac 12\psi _\varepsilon ^{+}\gamma \cdot \stackrel{\leftrightarrow
}{D}\psi _\varepsilon +{\rm i}m\psi _\varepsilon ^{+}\psi _\varepsilon
\rangle _c+\langle \psi _\varepsilon ^{+}\gamma _\mu F_{\mu \nu }\psi
_\varepsilon \rangle _c+{\cal A}_\nu ,
\end{eqnarray}
where the anomaly is given by the formula

\begin{equation}
{\cal A}_\nu =\langle \psi _\varepsilon ^{+}(\stackrel{\leftarrow }{D}_\nu [%
{\rm e}^{-\varepsilon \cdot \stackrel{\rightarrow }{D}},\gamma \cdot
\overrightarrow{D}]{\rm e}^{\varepsilon \cdot \stackrel{\rightarrow }{D}}-%
{\rm e}^{-\varepsilon \cdot \stackrel{\leftarrow }{D}}[\gamma \cdot
\stackrel{\leftarrow }{D},{\rm e}^{\varepsilon \cdot \stackrel{\leftarrow }{D%
}}]\stackrel{\rightarrow }{D}_\nu )\psi _\varepsilon \rangle _c.
\end{equation}
Expressing this in terms of the above introduced $G$ functions we have

\begin{eqnarray}
{\cal A}_\nu &=&\langle \psi _\varepsilon ^{+}(\stackrel{\leftarrow }{D}_\nu
\gamma \cdot G^{+}+\gamma \cdot G^{+}\stackrel{\rightarrow }{D}_\nu )\psi
_\varepsilon \rangle +\langle \psi _\varepsilon ^{+}(\stackrel{\leftarrow }{D%
}_\nu \gamma \cdot G^{-}-\gamma \cdot G^{-}\stackrel{\rightarrow }{D}_\nu
)\psi _\varepsilon \rangle  \nonumber \\
&=&\partial _\nu \langle \psi _\varepsilon ^{+}(\gamma \cdot G^{+})\psi
_\varepsilon \rangle -\langle \psi _\varepsilon ^{+}[D_\nu ,\gamma \cdot
G^{+}]\psi _\varepsilon \rangle  \nonumber \\
&&+\langle \psi _\varepsilon ^{+}(\stackrel{\leftarrow }{D}_\nu \gamma \cdot
G^{-}-\gamma \cdot G^{-}\stackrel{\rightarrow }{D}_\nu )\psi _\varepsilon
\rangle .
\end{eqnarray}
Let us work out also the last term. For the derivatives of the point-split
fields with respect to the parameter $\varepsilon $ we have the following
useful formulae

\begin{eqnarray}
\partial _\nu ^\varepsilon \psi _\varepsilon &=&-\int_0^1{\rm d}t\,{\rm e}%
^{-t\varepsilon \cdot D}D_\nu {\rm e}^{t\varepsilon \cdot D}\psi _\varepsilon
\nonumber \\
&=&-D_\nu \psi _\varepsilon +\Gamma _\nu (x,\varepsilon )\psi _\varepsilon
\label{DL}
\end{eqnarray}
and

\begin{eqnarray}
\partial _\nu ^\varepsilon \psi _\varepsilon ^{+} &=&\psi _\varepsilon
^{+}\int_0^1{\rm d}t\,{\rm e}^{-t\varepsilon \cdot \stackrel{\leftarrow }{D}}%
\stackrel{\leftarrow }{D}_\nu {\rm e}^{t\varepsilon \cdot \stackrel{%
\leftarrow }{D}}  \nonumber \\
\ &=&\psi _\varepsilon ^{+}\stackrel{\leftarrow }{D}_\nu +\psi _\varepsilon
^{+}\Gamma _\nu (x,-\varepsilon ),  \label{DR}
\end{eqnarray}
where $\partial ^\varepsilon $ means the partial derivative with respect to $%
\varepsilon $ and where $\Gamma _\mu (x,\varepsilon )=$ $H_\mu
(x,\varepsilon )-G_\mu (x,\varepsilon )$ (the functions $G_\mu
(x,\varepsilon )$ and $H_\mu (x,\varepsilon )$ were already introduced in
the previous sections, their explicit form\footnote{%
Note also, that we can also write
\begin{eqnarray*}
\Gamma _\mu (x,\varepsilon ) &=&-\int_0^1{\rm d}t[{\rm e}^{-t\varepsilon
\cdot D},D_\mu ]{\rm e}^{t\varepsilon \cdot D}=-\int_0^1{\rm d}tG_\mu
(x,t\varepsilon ) \\
\ &=&\sum_{n=0}^\infty \frac{(-1)^{n+1}}{(n+2)!}[\varepsilon \cdot
D,[\varepsilon \cdot D,[\ldots ,[\varepsilon \cdot D,\varepsilon _\nu F_{\mu
\nu }(x)]\ldots ]]]
\end{eqnarray*}
Here the second line is a consequence of the explicit formula for $G_\mu
(x,\varepsilon )$, which is given in Appendix A.} is given in Appendix A.)
I.e.

\begin{eqnarray}
{\cal A}_\nu &=&\partial _\nu \langle \psi _\varepsilon ^{+}(\gamma \cdot
G^{+})\psi _\varepsilon \rangle _c-\langle \psi _\varepsilon ^{+}[D_\nu
,\gamma \cdot G^{+}]\psi _\varepsilon \rangle _c  \nonumber \\
&&+\langle \psi _\varepsilon ^{+}(\overleftarrow{\partial _\nu ^\varepsilon }%
\gamma \cdot G^{-}+\gamma \cdot G^{-}\overrightarrow{\partial _\nu
^\varepsilon })\psi _\varepsilon \rangle _c  \nonumber \\
&&-\langle \psi _\varepsilon ^{+}(\Gamma _\nu (x,-\varepsilon )\gamma \cdot
G^{-}+\gamma \cdot G^{-}\Gamma _\nu (x,\varepsilon ))\psi _\varepsilon
\rangle _c,
\end{eqnarray}
or, in terms of the covariant propagator,
\begin{eqnarray}
{\cal A}_\nu &=&-{\rm Tr}(\partial _\nu ^\varepsilon S^{{\rm cov}})\gamma
\cdot G^{-}-\partial _\nu {\rm Tr}S^{{\rm cov}}\gamma \cdot G^{+}  \nonumber
\\
&&+{\rm Tr}S^{{\rm cov}}([D_\nu ,\gamma \cdot G^{+}]+\Gamma _\nu
(x,-\varepsilon )\gamma \cdot G^{-}+\gamma \cdot G^{-}\Gamma _\nu
(x,\varepsilon )).  \label{anomaly E-M}
\end{eqnarray}
Let us recall, that $G_\mu ^{+}(x,\varepsilon )={\cal O}(\varepsilon ^2)$
and $G_\mu ^{-}(x,\varepsilon )={\cal O}(\varepsilon ),$ $\Gamma _\mu
(x,\varepsilon )={\cal O}(\varepsilon )$, therefore in the second and third
term on the right hand side of (\ref{anomaly E-M}) only the first two terms
of the expansion of ${\rm Tr}S^{{\rm cov}}\gamma _\mu $ can contribute.
However, these terms are proportional to $\varepsilon _\mu $ and because $%
\varepsilon \cdot G^{\pm }(x,\varepsilon )=0$, their contribution vanishes.
So that only the first term on the right hand side of (\ref{anomaly E-M})
can survive and we get for the anomaly the following simple formula

\begin{equation}
{\cal A}_\nu =-{\rm Tr}(\partial _\nu ^\varepsilon S^{{\rm cov}})\gamma
\cdot G^{-}+{\cal O}(\varepsilon ).
\end{equation}
This gives, after some algebra\footnote{%
Note, that $G_\mu ^{-}=\varepsilon _\sigma F_{\mu \sigma }+{\cal O}%
(\varepsilon ^3)$ and
\begin{eqnarray*}
{\rm Tr}_DS^{{\rm cov}}\gamma _\mu &=&\frac 1{4\pi ^2}\left\{-\frac{\varepsilon
_\mu }{|\varepsilon |^4}+m^2\frac{\varepsilon _\mu }{|\varepsilon |^2}+\frac
13{\rm \ln }m^2|\varepsilon |^2[D_\sigma ,F_{\sigma \mu }]
\right.
\\
&&
\left.
\ \ -\frac 13\frac{\varepsilon _\mu \varepsilon _\sigma }{|\varepsilon |^2%
}[D_\rho ,F_{\rho \sigma }]+\frac 13\frac{\varepsilon _\rho \varepsilon
_\sigma }{|\varepsilon |^2}[D_\rho ,F_{\mu \sigma }]+{\cal O}_{{\rm reg}%
}(1)\right\},
\end{eqnarray*}
and therefore
\begin{eqnarray*}
{\rm Tr}_D\partial _\nu ^\varepsilon S^{{\rm cov}}\gamma _\mu &=&\frac
1{4\pi ^2}\left\{\frac{4\varepsilon _\mu \varepsilon _\nu -\delta _{\mu \nu
}|\varepsilon |^2}{|\varepsilon |^6}+m^2\frac{\delta _{\mu \nu }|\varepsilon
|^2-2\varepsilon _\mu \varepsilon _\nu }{|\varepsilon |^4}+\frac 23\frac{%
\varepsilon _\nu }{|\varepsilon |^2}[D_\sigma ,F_{\sigma \mu }]
\right.
\\
&&-\frac 13\frac{(\delta _{\mu \nu }\varepsilon _\sigma +\delta _{\nu \sigma
}\varepsilon _\mu )|\varepsilon |^2-2\varepsilon _\mu \varepsilon _\sigma
\varepsilon _\nu }{|\varepsilon |^4}[D_\rho ,F_{\rho \sigma }]
\\
&&
\left.
+\frac 13\frac{(\delta _{\nu \rho }\varepsilon _\sigma +\delta _{\nu
\sigma }\varepsilon _\rho )|\varepsilon |^2-2\varepsilon _\nu \varepsilon
_\rho \varepsilon _\sigma }{|\varepsilon |^4}[D_\rho ,F_{\mu \sigma }]+{\cal %
O}_{{\rm reg}}(\ln |\varepsilon |)\right\}.
\end{eqnarray*}
}

\begin{equation}
{\cal A}_\nu =-\frac 1{72\pi ^2}{\rm Tr}_C\left( 5[D_\alpha ,F_{\alpha \beta
}]F_{\beta \nu }+[D_\nu ,F_{\alpha \beta }]F_{\alpha \beta }+[D_\alpha
,F_{\beta \nu }]F_{\beta \alpha }\right) .
\end{equation}
Inserting here the identity $[D_\alpha ,F_{\beta \nu }]F_{\beta \alpha
}=\frac 12[D_\nu ,F_{\alpha \beta }]F_{\alpha \beta }$, we get finally

\begin{equation}
{\cal A}_\nu =-\frac 1{24\pi ^2}\partial _\mu {\rm Tr}_C\left( \frac
14\delta _{\mu \nu }F_{\alpha \beta }F_{\alpha \beta }+\frac 53\theta _{\mu
\nu }^A\right) +{\cal O}(\varepsilon )
\end{equation}
where
\begin{equation}
\theta _{\mu \nu }^A=F_{\mu \sigma }F_{\sigma \nu }+\frac 14\delta _{\mu \nu
}F_{\alpha \beta }F_{\alpha \beta }.
\end{equation}
As we have expected, the anomaly is shifted to the divergence of the
energy-momentum tensor. We can recover the trace anomaly by redefinition
(finite subtraction) of this tensor according to the prescription
\begin{equation}
\overline{\theta }_{\mu \nu }^{{\rm reg}}=\theta _{\mu \nu }^{{\rm reg}%
}+\frac 1{24\pi ^2}{\rm Tr}_C\left( \frac 14\delta _{\mu \nu }F_{\alpha
\beta }F_{\alpha \beta }+\frac 53\theta _{\mu \nu }^A\right) .
\end{equation}
The redefined energy-momentum tensor has the standard trace anomaly

\begin{equation}
\langle \overline{\theta }_{\mu \mu }^{{\rm reg}}\rangle _c=-{\rm i}\langle
m\psi _\varepsilon ^{+}\psi _\varepsilon \rangle _c+\frac 1{24\pi ^2}{\rm Tr}%
_CF_{\alpha \beta }F_{\alpha \beta }+{\cal O}(\varepsilon )
\end{equation}
and its divergence is anomaly free. This is an illustration of the anomalous
symmetry pair phenomenon.

Let us briefly comment on this result. As we have seen, the nonstandard
terms in the variation of the action under corresponding point-split
transformations are present both for the scale transformation and
translation. The anomalies are therefore not a priori excluded either in the
trace or in the divergence of the energy-momentum tensor. Although our point
of view does not explain why it appears in the divergence and not in the
trace (this result is hidden in the short distance asymptotics of the
covariant propagator), it at least sheds a new light on the well known fact
that point-splitting regularization is in a conflict with (covariant)
translations.


\section{The symmetric energy-momentum tensor and the Lorentz anomaly}


As we have mentioned above, the canonical energy-momentum tensor is not
symmetric. In this section we give construction of the modified
energy-momentum tensor, which does not suffer from this inconvenient
property.

Let us start with the local $SO(4)$ rotation of the fermion fields
\begin{eqnarray}
\delta \psi &=&\frac{{\rm i}}2\omega _{\mu \nu }\sigma _{\mu \nu }\psi
\nonumber \\
\delta \psi ^{+} &=&-\frac{{\rm i}}2\psi ^{+}\sigma _{\mu \nu }\omega _{\mu
\nu }  \label{o4rotation}
\end{eqnarray}
with the parameter $\omega _{\mu \nu }(x).$ The variation of the action with
respect to this transformation can be written as
\begin{equation}
\delta S_E=\frac{{\rm i}}4\int {\rm d}^4x\omega _{\mu \nu }\left[ -\partial
_\alpha (\psi ^{+}\{\gamma _\alpha ,\sigma _{\mu \nu }\}\psi )+4{\rm i}%
(\theta _{\mu \nu }-\theta _{\nu \mu })\right] .  \label{deltaSE}
\end{equation}
with $\theta _{\mu \nu }$ being the canonical energy-momentum tensor (\ref
{E-M}). Classically, using the identities
\begin{eqnarray}
\gamma _\alpha \sigma _{\mu \nu } &=&{\rm i}(\delta _{\nu \alpha }\gamma
_\mu -\delta _{\mu \alpha }\gamma _\nu +\varepsilon _{\mu \nu \alpha \lambda
}\gamma _\lambda \gamma _5)  \nonumber \\
\sigma _{\mu \nu }\gamma _\alpha &=&{\rm i}(\delta _{\mu \alpha }\gamma _\nu
-\delta _{\nu \alpha }\gamma _\mu +\varepsilon _{\mu \nu \alpha \lambda
}\gamma _\lambda \gamma _5).  \label{id}
\end{eqnarray}
we get then the following on-shell relation,
\begin{equation}
\theta _{\mu \nu }-\theta _{\nu \mu }=-\frac{{\rm i}}4\partial _\alpha (\psi
^{+}\{\gamma _\alpha ,\sigma _{\mu \nu }\}\psi )=\frac 12\varepsilon _{\mu
\nu \alpha \lambda }\partial _\alpha (\psi ^{+}\gamma _\lambda \gamma _5\psi
).  \label{antisymmetry}
\end{equation}
Thus, the classical symmetric energy-momentum tensor $T_{\mu \nu }$%
\begin{eqnarray}
T_{\mu \nu }=\frac 12(\theta _{\mu \nu }+\theta _{\nu \mu })=\frac 14(\psi
^{+}\gamma _\mu \stackrel{\leftrightarrow }{D}_\nu \psi +\psi ^{+}\gamma
_\nu \stackrel{\leftrightarrow }{D}_\mu \psi )
\end{eqnarray}
can be expressed on shell as
\begin{equation}
T_{\mu \nu }=\theta _{\mu \nu }-\frac 14\varepsilon _{\mu \nu \alpha \lambda
}\partial _\alpha (\psi ^{+}\gamma _\lambda \gamma _5\psi )  \label{TMN}
\end{equation}
and has therefore the same trace and divergence as the canonical tensor $%
\theta _{\mu \nu }.$

Let us now derive the quantum analog of the identity (\ref{TMN}), valid for
the point-split energy-momentum tensor$\ T_{\mu \nu }^{{\rm reg}}$ defined
as
\begin{equation}
T_{\mu \nu }^{{\rm reg}}=\frac 12(\theta _{\mu \nu }^{{\rm reg}}+\theta
_{\nu \mu }^{{\rm reg}})=\frac 14(\psi _\varepsilon ^{+}\gamma _\mu
\stackrel{\leftrightarrow }{D}_\nu \psi _\varepsilon +\psi _\varepsilon
^{+}\gamma _\nu \stackrel{\leftrightarrow }{D}_\mu \psi _\varepsilon )
\end{equation}

We shall use here again the idea of the point-split transformation. The
suitable deformation of the local $SO(4)$ rotation (\ref{o4rotation}) is
\begin{eqnarray}
\delta \psi &=&\frac{{\rm i}}2{\rm e}^{-\varepsilon \cdot D}\omega _{\mu \nu
}{\rm e}^{-\varepsilon \cdot D}\sigma _{\mu \nu }\psi  \nonumber \\
\delta \psi ^{+} &=&-\frac{{\rm i}}2\psi ^{+}\sigma _{\mu \nu }{\rm e}%
^{\varepsilon \cdot \stackrel{\leftarrow }{D}}\omega _{\mu \nu }{\rm e}%
^{\varepsilon \cdot \stackrel{\leftarrow }{D}}  \label{o4rotationreg}
\end{eqnarray}
The variation of the action under this transformation gives the following
regularized form of the formula (\ref{deltaSE}) and
\begin{eqnarray}
\delta S_E &=&\frac{{\rm i}}4\int {\rm d}^4x\omega _{\mu \nu }[-\partial
_\alpha (\psi _\varepsilon ^{+}\{\gamma _\alpha ,\sigma _{\mu \nu }\}\psi
_\varepsilon )
+4{\rm i}(\theta _{\mu \nu }^{{\rm reg}}-\theta _{\nu \mu }^{%
{\rm reg}})
\nonumber  \label{deltaSE2}
\\
&&\ \ \ \ -2\psi _\varepsilon ^{+}(\sigma _{\mu \nu }[{\rm e}^{-\varepsilon
\cdot \stackrel{\rightarrow }{D}},\gamma \cdot \stackrel{\rightarrow }{D}]{\rm e}%
^{\varepsilon \cdot \stackrel{\rightarrow }{D}}+{\rm e}^{-\varepsilon \cdot
\stackrel{\leftarrow }{D}}[\gamma \cdot \stackrel{\leftarrow }{D},{\rm e}%
^{\varepsilon \cdot \stackrel{\leftarrow }{D}}]\sigma _{\mu \nu })\psi
_\varepsilon ].  \label{deltaSEreg}
\end{eqnarray}
The Ward-Takahashi identity corresponding to the transformation (\ref
{o4rotationreg}) can be therefore rewritten in the form
\begin{equation}
\langle \theta _{\mu \nu }^{{\rm reg}}-\theta _{\nu \mu }^{{\rm reg}}\rangle
_c=\frac 12\varepsilon _{\mu \nu \alpha \lambda }\partial _\alpha \langle
\psi _\varepsilon ^{+}\gamma _\lambda \gamma _5\psi _\varepsilon \rangle _c+%
{\cal A}_{\mu \nu }
\end{equation}
where
\begin{equation}
{\cal A}_{\mu \nu }=-{\cal A}_{\nu \mu }=-\frac{{\rm i}}2\langle \psi
_\varepsilon ^{+}(\sigma _{\mu \nu }[{\rm e}^{-\varepsilon \cdot
\stackrel{\rightarrow }{D}},\gamma \cdot \stackrel{\rightarrow }{D}]{\rm e}%
^{\varepsilon \cdot \stackrel{\rightarrow }{D}}+{\rm e}^{-\varepsilon \cdot
\stackrel{\leftarrow }{D}}[\gamma \cdot \stackrel{\leftarrow }{D},{\rm e}%
^{\varepsilon \cdot \stackrel{\leftarrow }{D}}]\sigma _{\mu \nu })\psi
_\varepsilon \rangle _c  \label{o4anomaly}
\end{equation}
is a possible anomaly term corresponding to the violation of the naive
identity (\ref{antisymmetry}). Such an anomaly is called Lorentz anomaly.
The regularized form of the symmetric energy-momentum tensor $T_{\mu \nu }^{%
{\rm reg}}$ is therefore\footnote{%
This operator identity should be understood as the statement for the
corresponding generating functionals with operator insertions, i.e.
\[
T_{\mu \nu }^{{\rm reg}}\cdot Z_E[A]_c=\theta _{\mu \nu }^{{\rm reg}}\cdot
Z_E[A]_c-\frac 14\varepsilon _{\mu \nu \alpha \lambda }\partial _\alpha
(\psi _\varepsilon ^{+}\gamma _\lambda \gamma _5\psi _\varepsilon )\cdot
Z_E[A]_c-\frac 12{\cal A}_{\mu \nu }.
\]
}
\begin{equation}
T_{\mu \nu }^{{\rm reg}}=\theta _{\mu \nu }^{{\rm reg}}-\frac 14\varepsilon
_{\mu \nu \alpha \lambda }\partial _\alpha (\psi _\varepsilon ^{+}\gamma
_\lambda \gamma _5\psi _\varepsilon )-\frac 12{\cal A}_{\mu \nu }.
\label{tmn}
\end{equation}
Provided the anomaly does not vanish, we may get modification of the
divergence of the quantum symmetric energy-momentum $T_{\mu \nu }^{{\rm reg}%
} $ tensor, which, unlike in the classical case, would not be the same as
the divergence of the canonical tensor $\theta _{\mu \nu }^{{\rm reg}}$. For
the anomaly we have, using the identities (\ref{id}) and the definition of
the functions $G(x,\varepsilon )$ and $G^{\pm }(x,\varepsilon )$
\begin{eqnarray}
{\cal A}_{\mu \nu } &=&\frac{{\rm i}}2{\rm Tr}S^{{\rm cov}}(\sigma _{\mu \nu
}\gamma \cdot G(x,\varepsilon )-G(x,-\varepsilon )\cdot \gamma \sigma _{\mu
\nu })  \nonumber \\
\ &=&-{\rm Tr}S^{{\rm cov}}[(\delta _{\mu \alpha }\gamma _\nu -\delta _{\nu
\alpha }\gamma _\mu )G_\alpha ^{+}(x,\varepsilon )+\varepsilon _{\mu \nu
\alpha \lambda }\gamma _\lambda \gamma _5G_\alpha ^{-}(x,\varepsilon )].
\end{eqnarray}
Using now the explicit formulae for $S^{{\rm cov}}$ and the functions $%
G(x,\varepsilon )$ and $G^{\pm }(x,\varepsilon ),$ it is easy to show, that
in the symmetric limit $\varepsilon \rightarrow 0$ we get
\begin{equation}
{\cal A}_{\mu \nu }={\cal O}(\varepsilon )
\end{equation}
i.e., there is no $SO(4)$ anomaly and the regularized symmetric
energy-momentum tensor (\ref{tmn}) has the same trace and divergence as the
canonical one in the limit of the removed cut-off.


\section{Non-covariant point-splitting}


In the previous sections we have introduced a gauge covariant version of the
point-splitting regularization for various Noether currents. In this section
we relax this restriction and try to show how the violation of the gauge
covariance effects the structure of the anomalies discussed above.

The simplest type of the non-covariant point-splitting can be achieved by
replacing the parallel transporter $\Omega (x,y)$ (which ensured the gauge
covariance of the regularized currents) by a deformed one according to the
prescription
\begin{eqnarray}
\Omega (x,y) &=&P\exp \left( -\int_0^1{\rm d}t(x-y)\cdot A(y+t(x-y))\right)
\rightarrow  \nonumber \\
\ &\rightarrow &P\exp \left( -a\int_0^1{\rm d}t(x-y)\cdot A(y+t(x-y))\right)
=\widetilde{\Omega }(x,y)
\end{eqnarray}
where $a$ is a real parameter\footnote{%
That means $\widetilde{\Omega }(x,y)$ corresponds to the paralell
transporter allong the straight line connecting the points $x$ and $y$ in
the gauge field $aA(x)$ instead of $A(x)$.}. Gauge covariance is then
recovered putting $a=1$. For further convenience let us introduce the
following notation
\begin{eqnarray}
\widetilde{\psi }_\varepsilon (x) &=&\widetilde{\Omega }(x,x-\varepsilon
)\psi (x-\varepsilon )={\rm e}^{-\varepsilon \cdot \widetilde{D}}\psi
(x)=\Omega _{-}\psi _\varepsilon (x),  \nonumber \\
\widetilde{\psi }_\varepsilon ^{+}(x) &=&\psi ^{+}(x+\varepsilon )\widetilde{%
\Omega }(x+\varepsilon ,x)=\psi ^{+}(x){\rm e}^{\varepsilon \cdot \stackrel{%
\leftarrow }{\widetilde{D}}}=\psi _\varepsilon ^{+}(x)\Omega _{+}.
\end{eqnarray}
Here $\widetilde{D}=\partial +aA$ and, using the results of the previous
sections,
\begin{eqnarray}
\Omega _{-} &=&\widetilde{\Omega }(x,x-\varepsilon )\Omega
^{+}(x,x-\varepsilon )={\rm e}^{-\varepsilon \cdot \widetilde{D}}{\rm e}%
^{\varepsilon \cdot D},  \nonumber \\
\Omega _{+} &=&\Omega ^{+}(x+\varepsilon ,x)\widetilde{\Omega }%
(x+\varepsilon ,x)={\rm e}^{-\varepsilon \cdot \stackrel{\leftarrow }{D}}%
{\rm e}^{\varepsilon \cdot \stackrel{\leftarrow }{\widetilde{D}}}.
\end{eqnarray}
From the operator expressions the $\varepsilon -$expansion of these matrix
functions can be easily obtained, the explicit formulae are given in
Appendix B. The non-covariantly regularized currents are now built from the
modified point-split fields $\widetilde{\psi }_\varepsilon (x)$ and $%
\widetilde{\psi }_\varepsilon ^{+}(x)$ and (in analogy with the previous
sections) they can be understood as the Noether currents corresponding to
the modified point-split transformations (\ref{deformed chiral rotation}), (%
\ref{deformed gauge rotation}), (\ref{deformed gauge rotation}) and (\ref
{deformed translation}). Such modification is given by replacing the
covariant derivative with the deformed one $D\rightarrow \widetilde{D}$ e.g.
the modified point-split chiral rotation reads
\begin{eqnarray}
\widetilde{\delta }\psi &=&{\rm e}^{-\varepsilon \cdot \widetilde{D}}\alpha
\gamma _5{\rm e}^{-\varepsilon \cdot \widetilde{D}}\psi ,  \nonumber \\
\widetilde{\delta }\psi ^{+} &=&\psi ^{+}{\rm e}^{\varepsilon \cdot
\stackrel{\leftarrow }{\widetilde{D}}}\alpha \gamma _5{\rm e}^{\varepsilon
\cdot \stackrel{\leftarrow }{\widetilde{D}}}.
\label{modified deformed chiral rotation}
\end{eqnarray}

The above described change of the regularization scheme modifies generally
the Noether currents $j_\mu ^a$, $j_{\mu 5}^a$, $\theta _{\mu \nu }$ and the
composite operators $s^a=\psi ^{+}T^a\psi $ and $p^a=\psi ^{+}T^a\gamma
_5\psi $ entering the Ward-Takahashi identities by means of adding
non-covariant counterterms $\Delta \widetilde{j}_\mu ^a$ , $\Delta
\widetilde{j}_{\mu 5}^a,$ $\Delta \widetilde{\theta }_{\mu \nu }$, $\Delta
\widetilde{s}^a$ and $\Delta \widetilde{p}^a$ respectively. In this section
we give a list of such counterterms and deduce from it the additional
spurious contributions to the anomalies. These contributions can be
expressed by the following formulae, which follow from the regularized form
of the Ward-Takahashi identities\footnote{%
Here $j_\mu =\psi ^{+}\gamma _\mu \psi $ and $j_{\mu 5}=\psi ^{+}\gamma _\mu
\gamma _5\psi $ are the singlet currents and $\Delta \widetilde{j}_\mu $ and
$\Delta \widetilde{j}_{\mu 5}$ are corresponding counterterms. The same
notation without superscript is used for the singlet scalar and pseudoscalar
densities.}
\begin{eqnarray}
\Delta \widetilde{{\cal A}}^a &=&D_\mu \Delta \widetilde{j}_\mu ^a  \nonumber
\\
\Delta \widetilde{{\cal A}}_5^a &=&D_\mu \Delta \widetilde{j}_{\mu 5}^a-2%
{\rm i}m\Delta \widetilde{p}^a  \nonumber \\
\Delta \widetilde{{\cal A}}^{{\rm trace}} &=&\Delta \widetilde{\theta }_{\mu
\mu }+{\rm i}m\Delta \widetilde{s}  \nonumber \\
\Delta \widetilde{{\cal A}}_\nu &=&\partial _\mu \Delta \widetilde{\theta }%
_{\mu \nu }-\partial _\nu (\Delta \widetilde{\theta }_{\mu \mu }+{\rm i}%
m\Delta \widetilde{s})-F_{\mu \nu }^a\Delta \widetilde{j}_\mu ^a  \nonumber
\\
\Delta \widetilde{{\cal A}}_{\mu \nu } &=&\Delta \widetilde{\theta }_{\mu
\nu }-\Delta \widetilde{\theta }_{\nu \mu }-\frac 12\varepsilon _{\mu \nu
\alpha \lambda }\partial _\alpha \Delta \widetilde{j}_{\lambda 5}
\end{eqnarray}

Let us start with the axial currents. The non-covariant point-split
regularization of this operator is
\begin{eqnarray}
\widetilde{j}_{5\mu }^{a,{\rm reg}}(x) &=&\psi ^{+}(x+\varepsilon )%
\widetilde{\Omega }(x+\varepsilon ,x)T^a\gamma _\mu \gamma _5\widetilde{%
\Omega }(x,x-\varepsilon )\psi (x-\varepsilon )  \nonumber \\
\ &=&\widetilde{\psi }_\varepsilon ^{+}T^a\gamma _\mu \gamma _5\widetilde{%
\psi }_\varepsilon =\psi _\varepsilon ^{+}\Omega _{+}T^a\gamma _\mu \gamma
_5\Omega _{-}\psi _\varepsilon .
\end{eqnarray}
and it differs from the covariant one by the following counterterm
\begin{eqnarray}
\Delta \widetilde{j}_{5\mu }^{a,{\rm reg}} &=&\langle \widetilde{\psi }%
_\varepsilon ^{+}T^a\gamma _\mu \gamma _5\widetilde{\psi }_\varepsilon -\psi
_\varepsilon ^{+}T^a\gamma _\mu \gamma _5\psi _\varepsilon \rangle  \nonumber
\\
\ &=&-{\rm Tr}S^{{\rm cov}}\gamma _\mu \gamma _5(\Omega _{+}T^a\Omega
_{-}-T^a).
\end{eqnarray}
Using the expansion
\begin{equation}
\Omega _{\pm }=1+(1-a)\varepsilon \cdot A(x)+{\cal O}(\varepsilon ^2)
\end{equation}
we get
\begin{equation}
\Delta \widetilde{j}_{5\mu }^{a,{\rm reg}}=\frac{1-a}{4\pi ^2}{\rm Tr}%
_C\left( \left(\frac{\varepsilon _\nu }{|\varepsilon |^2}F_{\nu \mu }^{*}(x)+%
{\cal O}(1)\right)(\{T^a,\varepsilon \cdot A(x)\}+{\cal O}(\varepsilon ^2))\right)
\end{equation}
and after the symmetrization over the direction of $\varepsilon $%
\begin{equation}
\Delta \widetilde{j}_{5\mu }^{a,{\rm reg}}=\frac{1-a}{16\pi ^2}{\rm Tr}%
_CT^a\{A_\nu (x),F_{\nu \mu }^{*}(x)\}+{\cal O}(\varepsilon ).
\end{equation}
In the same way we can prove, that for the pseudoscalar density $p=\psi
^{+}\gamma _5\psi $ we get counterterm which vanishes in the limit of the
removed point-splitting
\begin{eqnarray}
\Delta \widetilde{p}^{a,{\rm reg}} &=&\langle \widetilde{\psi }_\varepsilon
^{+}T^a\gamma _5\widetilde{\psi }_\varepsilon -\psi _\varepsilon
^{+}T^a\gamma _5\psi _\varepsilon \rangle =-{\rm Tr}S^{{\rm cov}}\gamma
_5(\Omega _{+}T^a\Omega _{-}-T^a)  \nonumber \\
\ &=&{\cal O}(\varepsilon ).
\end{eqnarray}
The chiral Ward-Takahashi identity has now the following form
\begin{equation}
D_\mu \langle \widetilde{j}_{5\mu }^{a,{\rm reg}}\rangle =2{\rm i}m\langle
\widetilde{\psi }_\varepsilon ^{+}T^a\gamma _5\widetilde{\psi }_\varepsilon
\rangle +\widetilde{{\cal A}}_5^a
\end{equation}
where
\begin{equation}
\widetilde{{\cal A}}_5^a={\cal A}_5^a+D_\mu \Delta \widetilde{j}_{5\mu }^{a,%
{\rm reg}}=\frac 1{8\pi ^2}{\rm Tr}_CT^aF_{\mu \nu }^{*}(x)F_{\mu \nu }(x)+%
\frac{1-a}{16\pi ^2}{\rm Tr}_CT^a\{[D_\mu ,A_\nu (x)],F_{\nu \mu }^{*}(x)\}+%
{\cal O}(\varepsilon ).  \label{chiral tilde}
\end{equation}
Note, that for the abelian case we have
\begin{equation}
\widetilde{{\cal A}}_5=\frac{1+a}{16\pi ^2}F_{\mu \nu }^{*}(x)F_{\mu \nu }(x)
\label{abelian5}
\end{equation}
and the anomaly can be eliminated taking $a=-1$, cf. \cite{Urrutia}.

Because the gauge symmetry is violated in this regularization scheme, we
expect also anomalous divergence of the gauge current
\begin{equation}
\widetilde{j}_\mu ^{a,{\rm reg}}=\widetilde{\psi }_\varepsilon ^{+}T^a\gamma
_\mu \widetilde{\psi }_\varepsilon =\psi _\varepsilon ^{+}\Omega
_{+}T^a\gamma _\mu \Omega _{-}\psi _\varepsilon .  \label{jnoncov}
\end{equation}
For the counterterm we have
\begin{eqnarray}
\Delta \widetilde{j}_\mu ^{a,{\rm reg}} &=&\langle \widetilde{\psi }%
_\varepsilon ^{+}T^a\gamma _\mu \widetilde{\psi }_\varepsilon -\psi
_\varepsilon ^{+}T^a\gamma _\mu \psi _\varepsilon \rangle  \nonumber \\
\ &=&-{\rm Tr}S^{{\rm cov}}\gamma _\mu (\Omega _{+}T^a\Omega _{-}-T^a)
\nonumber \\
\ &=&\frac 1{4\pi ^2}\left( \frac{\varepsilon _\mu }{|\varepsilon |^4}-m^2%
\frac{\varepsilon _\mu }{|\varepsilon |^2}+{\cal O}(\ln |\varepsilon
|)\right) {\rm Tr}_C(\Omega _{+}T^a\Omega _{-}-T^a)  \label{jctnoncov}
\end{eqnarray}
Using the formula (see Appendix B)
\begin{eqnarray}
\Omega _{-}\Omega _{+} &=&1+2(1-a)(A(x)\cdot \varepsilon
)+2(1-a)^2(A(x)\cdot \varepsilon )^2  \nonumber \\
&&\ \ \ \ \ +\frac 43(1-a)^3(A(x)\cdot \varepsilon )^3+\frac
13(1-a)(2-a)[(A(x)\cdot \varepsilon ),(\varepsilon \cdot \partial
)(A(x)\cdot \varepsilon )]  \nonumber \\
&&\ \ \ \ \ +\frac 13(1-a)(\varepsilon \cdot \partial )^2(A(x)\cdot
\varepsilon )+{\cal O}(\varepsilon ^4)  \label{omega-omega+}
\end{eqnarray}
we get after the symmetrization over the direction of $\varepsilon $%
\begin{eqnarray}
\Delta \widetilde{j}_\mu ^{a,{\rm reg}} &=&\frac{1-a}{8\pi ^2}{\rm Tr}%
_CT^a\left((\frac 1{|\varepsilon |^2}-m^2)A_\mu
\right.
\nonumber \\
&&\ \ \ \ \ +\frac{(1-a)^2}9(A_\mu A\cdot A+A\cdot AA_\mu +A_\nu A_\mu A_\nu
)
\nonumber \\
&&\ \ \ \ \ +\frac{(2-a)}{36}([A_\mu ,\partial \cdot A]+[A_\nu ,\partial
_\mu A_\nu ]+[A_\nu ,\partial _\nu A_\mu ])
\nonumber \\
&&
\left.
\ \ \ \ \ +\frac 1{36}(2\partial _\mu \partial \cdot A+\partial ^2A_\mu )
\right)+%
{\cal O}(\varepsilon ).
\end{eqnarray}
The regularized current $\widetilde{j}_\mu ^{a,{\rm reg}}$ has therefore
anomalous divergence
\begin{equation}
D_\mu \langle \widetilde{j}_\mu ^{a,{\rm reg}}\rangle =\widetilde{{\cal A}}^a
\end{equation}
and the gauge anomaly is
\begin{eqnarray}
\widetilde{{\cal A}}^a &=&\frac{1-a}{8\pi ^2}{\rm Tr}_CT^a[D_\mu ,
((\frac 1{|\varepsilon |^2}-m^2)A_\mu
\nonumber \\
&&\ \ \ \ \ +\frac{(1-a)^2}9(A_\mu A\cdot A+A\cdot AA_\mu +A_\nu A_\mu A_\nu
)
\nonumber \\
&&\ \ \ \ \ +\frac{(2-a)}{36}([A_\mu ,\partial \cdot A]+[A_\nu ,\partial
_\mu A_\nu ]+[A_\nu ,\partial _\nu A_\mu ])
\nonumber \\
&&
\ \ \ \ \ +\frac 1{36}(2\partial _\mu \partial \cdot A+\partial ^2A_\mu
))]+{\cal O}(\varepsilon ).  \label{vector tilde}
\end{eqnarray}
In the abelian case (cf. \cite{Urrutia}) we get
\begin{equation}
\widetilde{{\cal A}}=\frac{1-a}{8\pi ^2}\left(\left(\frac 1{|\varepsilon
|^2}-m^2\right)\partial \cdot A+\frac 1{12}\partial ^2\partial \cdot A+\frac{%
(1-a)^2}3(A^2\partial \cdot A+2A_\nu A\cdot \partial A_\nu)
\right)
\label{abelian}
\end{equation}
which together with (\ref{abelian5}) reproduces the well known one-parameter
family of gauge and chiral anomalies \cite{Urrutia}.

Let us now calculate the non-covariantly regularized energy-momentum tensor.
We have
\begin{equation}
\widetilde{\theta }_{\mu \nu }^{{\rm reg}}=\frac 12\widetilde{\psi }%
_\varepsilon ^{+}\gamma _\mu \stackrel{\leftrightarrow }{D_\nu }\widetilde{%
\psi }_\varepsilon =\frac 12\psi _\varepsilon ^{+}\Omega _{+}\gamma _\mu
\stackrel{\leftrightarrow }{D_\nu }\Omega _{-}\psi _\varepsilon
\end{equation}
and the non-covariant counterterm can be rewritten after simple manipulation
in the form
\begin{equation}
\Delta \widetilde{\theta }_{\mu \nu }^{{\rm reg}}=\frac 12\langle \psi
_\varepsilon ^{+}\gamma _\mu (\Omega _{+}[\stackrel{\rightarrow }{D}_\nu
,\Omega _{-}]-[\Omega _{+},\stackrel{\leftarrow }{D}_\nu ]\Omega
_{-}+(\Omega _{+}\Omega _{-}-1)\stackrel{\rightarrow }{D}_\nu -\stackrel{%
\leftarrow }{D}_\nu (\Omega _{+}\Omega _{-}-1))\psi _\varepsilon \rangle .
\end{equation}
Using the identities (\ref{DL}) and (\ref{DR}), we have then
\begin{eqnarray}
\Delta \widetilde{\theta }_{\mu \nu }^{{\rm reg}} &=&-\frac 12{\rm Tr}S^{%
{\rm cov}}\gamma _\mu (\Omega _{+}[\stackrel{\rightarrow }{D}_\nu ,\Omega
_{-}]-[\Omega _{+},\stackrel{\leftarrow }{D}_\nu ]\Omega _{-}  \nonumber \\
&&\ \ \ \ +(\Omega _{+}\Omega _{-}-1)\Gamma _\nu (x,\varepsilon )+\Gamma
_\nu (x,-\varepsilon )(\Omega _{+}\Omega _{-}-1))  \nonumber \\
&&\ \ \ \ \ \ +\frac 12{\rm Tr}(\partial _\nu ^\varepsilon S^{{\rm cov}%
}\gamma _\mu )(\Omega _{+}\Omega _{-}-1).
\end{eqnarray}
After some algebra (the details are given in Appendix B) we get
\begin{eqnarray}
\Delta \widetilde{\theta }_{\mu \nu }^{{\rm reg}}
&=&\frac{(1-a)^2}{48\pi ^2}%
{\rm Tr}_C\left(\delta _{\mu \nu }A^2\left(2m^2-\frac 1{|\varepsilon
|^2}\right)-2A_\mu A_\nu \left(m^2-\frac 2{|\varepsilon
|^2}\right)\right)
\nonumber \\
&&\ \ \ \ -\frac{(1-a)}{576\pi ^2}{\rm Tr}_C(8\delta _{\mu \nu }A_\alpha
[D_\beta ,F_{\beta \alpha }]-11(A_\mu [D_\alpha ,F_{\alpha \nu }]+A_\nu
[D_\alpha ,F_{\alpha \mu }])
\nonumber \\
&&\ \ \ \ -5(A_\alpha [D_\nu ,F_{\mu \alpha }]+A_\alpha [D_\mu ,F_{\nu
\alpha }])  \nonumber \\
&&\ \ \ \ +(1-a)(\delta _{\mu \nu }(2A\cdot \partial \partial \cdot
A+A_\alpha \partial ^2A_\alpha )+(A_\mu \partial _\nu +A_\nu \partial _\mu
)\partial \cdot A  \nonumber \\
&&\ \ \ \ +A\cdot \partial (\partial _\nu A_\mu +\partial _\mu A_\nu
)+4A_\alpha \partial _\mu \partial _\nu A_\alpha -A_\mu \partial ^2A_\nu
-A_\nu \partial ^2A_\mu  \nonumber \\
&&\ \ \ \ +3([A_\alpha ,A_\mu ](\partial _\nu A_\alpha +\partial _\alpha
A_\nu )+[A_\alpha ,A_\nu ](\partial _\mu A_\alpha +\partial _\alpha A_\mu )))
\nonumber \\
&&\ \ \ \ +(1-a)^2(\delta _{\mu \nu }(2A^4+A_\alpha A_\beta A_\alpha A_\beta
)-4(\{A_\mu ,A_\nu \}A^2+A_\mu A_\alpha A_\nu A_\alpha )))  \nonumber \\
&&\ \ \ \ -\frac{(1-a)}{192\pi ^2}{\rm Tr}_C((a+6)(A_\mu [D_\alpha
,F_{\alpha \nu }]-A_\nu [D_\alpha ,F_{\alpha \mu }])+(a-4)A\cdot \partial
F_{\mu \nu }  \nonumber \\
&&\ \ \ \ +(1-a)(A_\mu \partial ^2A_\nu -A_\nu \partial ^2A_\mu -2A\cdot
\partial [A_\mu ,A_\nu ]))  \label{counterterm theta}
\end{eqnarray}
where we picked up the divergent, symmetric and antisymmetric part.

Let us calculate also the non-covariant counterterm for the scalar density $%
s=\psi ^{+}\psi $, we have using (\ref{omega-omega+})
\begin{eqnarray}
\Delta \widetilde{s} &=&\langle \widetilde{\psi }_\varepsilon ^{+}\widetilde{%
\psi }_\varepsilon -\psi _\varepsilon ^{+}\psi _\varepsilon \rangle =-{\rm Tr%
}S^{{\rm cov}}(\Omega _{+}\Omega _{-}-1)  \nonumber \\
\ &=&\frac{(1-a)^2}{8\pi ^2}{\rm i}m{\rm Tr}_CA^2.
\end{eqnarray}
For the trace anomaly in the non-covariant regularization scheme we get then
\begin{eqnarray}
\widetilde{{\cal A}}^{{\rm trace}}=\Delta \widetilde{\theta }_{\mu \mu }^{%
{\rm reg}}+{\rm i}m\Delta \widetilde{s}=-\frac{(1-a)^2}{96\pi ^2}{\rm Tr}%
_C(2A\cdot \partial \partial \cdot A+A_\alpha \partial ^2A_\alpha ),
\label{tracetilde}
\end{eqnarray}
The spurious Lorentz anomaly can be also easily found as
\begin{eqnarray}
\widetilde{{\cal A}}_{\mu \nu } &=&\Delta \widetilde{\theta }_{\mu \nu }^{%
{\rm reg}}-\Delta \widetilde{\theta }_{\nu \mu }^{{\rm reg}}-\frac
12\varepsilon _{\mu \nu \alpha \lambda }\partial _\alpha \Delta \widetilde{j}%
_{5\lambda }^{{\rm reg}}
\nonumber \\
&=& -\frac{(1-a)}{96\pi ^2}{\rm Tr}_C((a+6)(A_\mu [D_\alpha ,F_{\alpha \nu
}]-A_\nu [D_\alpha ,F_{\alpha \mu }])+(a-4)A\cdot \partial F_{\mu \nu }
\nonumber \\
&& +(1-a)(A_\mu \partial ^2A_\nu -A_\nu \partial ^2A_\mu -2A\cdot
\partial [A_\mu ,A_\nu ]))  \nonumber \\
&& -\frac{1-a}{16\pi ^2}\partial _\alpha {\rm Tr}_C(A_\alpha F_{\mu \nu
}+A_\mu F_{\nu \alpha }-A_\nu F_{\mu \alpha }),
\end{eqnarray}
or, after some algebra,
\begin{eqnarray}
\widetilde{{\cal A}}_{\mu \nu } &=&-\frac{1-a}{16\pi ^2}(a(A_\mu [D_\alpha
,F_{\alpha \nu }]-A_\nu [D_\alpha ,F_{\alpha \mu }])+(a+2)A\cdot \partial
F_{\mu \nu }  \nonumber \\
&&\ \ \ +6([D_\alpha ,A_\mu ]\partial _\nu A_\alpha -[D_\alpha ,A_\nu
]\partial _\mu A_\alpha +F_{\mu \nu }\partial \cdot A)  \nonumber \\
&&\ \ \ +(1-a)(A_\mu \partial ^2A_\nu -A_\nu \partial ^2A_\mu -2A\cdot
\partial [A_\mu ,A_\nu ])).  \label{so(4)tilde}
\end{eqnarray}
Especially in the abelian case we have
\begin{eqnarray}
\widetilde{{\cal A}}_{\mu \nu }\
&=&\frac{1-a}{16\pi ^2}\biggl(\frac a6(A_\mu
\partial _\nu -A_\nu \partial _\mu )\partial \cdot A-\frac 16(2+a)A\cdot
\partial F_{\mu \nu }
\nonumber \\
&&\ \ \ \ -\frac 16(A_\mu \partial ^2A_\nu -A_\nu \partial ^2A_\mu )
\nonumber \\
&&\ \ \ \ \ \ \ -F_{\mu \nu }\partial \cdot A-\ (\partial _\alpha A_\mu
)(\partial _\nu A_\alpha )+(\partial _\alpha A_\nu )(\partial _\mu A_\alpha
)\biggr).
\end{eqnarray}
For the translation anomaly we get
\begin{eqnarray}
\Delta \widetilde{{\cal A}}_\nu
&=&\partial _\mu \Delta \widetilde{\theta }%
_{\mu \nu }-\partial _\nu \widetilde{{\cal A}}^{{\rm trace}}-F_{\mu \nu
}^a\Delta \widetilde{j}_\mu ^a
\nonumber \\
&=&\frac{(1-a)^2}{48\pi ^2}{\rm Tr}_C
\left(\partial _\nu A^2\left(2m^2-\frac 1{|\varepsilon |^2}\right)
-2\partial _\mu (A_\mu A_\nu )\left(m^2-\frac 2{|\varepsilon|^2}\right)
\right.
\nonumber \\
&&
\left. \ \ \ \ \
-6F_{\mu \nu }A_\mu \left(\frac 1{|\varepsilon|^2}-m^2\right)\right)
\nonumber
\\
&&-\frac{(1-a)}{576\pi ^2}\partial _\mu {\rm Tr}_C(8\delta _{\mu \nu
}A_\alpha [D_\beta ,F_{\beta \alpha }]-11(A_\mu [D_\alpha ,F_{\alpha \nu
}]+A_\nu [D_\alpha ,F_{\alpha \mu }])  \nonumber \\
&&\ \ \ \ \ -5(A_\alpha [D_\nu ,F_{\mu \alpha }]+A_\alpha [D_\mu ,F_{\nu
\alpha }])  \nonumber \\
&&\ \ \ \ \ +(1-a)(-5\delta _{\mu \nu }(2A\cdot \partial \partial \cdot
A+A_\alpha \partial ^2A_\alpha )+(A_\mu \partial _\nu +A_\nu \partial _\mu
)\partial \cdot A  \nonumber \\
&&\ \ \ \ \ +A\cdot \partial (\partial _\nu A_\mu +\partial _\mu A_\nu
)+4A_\alpha \partial _\mu \partial _\nu A_\alpha -A_\mu \partial ^2A_\nu
-A_\nu \partial ^2A_\mu  \nonumber \\
&&\ \ \ \ \ +3([A_\alpha ,A_\mu ](\partial _\nu A_\alpha +\partial _\alpha
A_\nu )+[A_\alpha ,A_\nu ](\partial _\mu A_\alpha +\partial _\alpha A_\mu )))
\nonumber \\
&&\ \ \ \ \ +(1-a)^2(\delta _{\mu \nu }(2A^4+A_\alpha A_\beta A_\alpha
A_\beta )-4(\{A_\mu ,A_\nu \}A^2+A_\mu A_\alpha A_\nu A_\alpha )))  \nonumber
\\
&&-\frac{(1-a)}{192\pi ^2}\partial _\mu {\rm Tr}_C((a+6)(A_\mu [D_\alpha
,F_{\alpha \nu }]-A_\nu [D_\alpha ,F_{\alpha \mu }])+(a-4)A\cdot \partial
F_{\mu \nu }  \nonumber \\
&&\ \ \ \ \ +(1-a)(A_\mu \partial ^2A_\nu -A_\nu \partial ^2A_\mu -2A\cdot
\partial [A_\mu ,A_\nu ]))  \nonumber \\
&&-\frac{(1-a)}{288\pi ^2}{\rm Tr}_CF_{\mu \nu }(4(1-a)^2(A_\mu A\cdot
A+A\cdot AA_\mu +A_\nu A_\mu A_\nu )  \nonumber \\
&&\ \ \ \ \ +(2-a)([A_\mu ,\partial \cdot A]+[A_\nu ,\partial _\mu A_\nu
]+[A_\nu ,\partial _\nu A_\mu ])  \nonumber \\
&&\ \ \ \ \ +(2\partial _\mu \partial \cdot A+\partial ^2A_\mu )))
\label{translationtilde}
\end{eqnarray}

Let us note, that the formulae (\ref{chiral tilde}), (\ref{vector tilde}), (%
\ref{tracetilde}), (\ref{so(4)tilde}) and (\ref{translationtilde}) for the
anomalies can be also obtained by means of the corresponding modification of
the Ward-Takahashi identities approach used in the previous sections. E.g.
we have the following Ward-Takahashi identity for the non-covariantly
point-split chiral transformation (\ref{modified deformed chiral rotation})
\begin{eqnarray}
\delta S_E &=&\int {\rm d}^4x\left( \widetilde{\psi }_\varepsilon
^{+}[\gamma \cdot D,\alpha ]\gamma _5\widetilde{\psi }_\varepsilon +2{\rm i}m%
\widetilde{\psi }_\varepsilon ^{+}\alpha \gamma _5\widetilde{\psi }%
_\varepsilon \right)  \nonumber \\
&&\ \ +\int {\rm d}^4x\left( \widetilde{\psi }_\varepsilon ^{+}(\alpha
\gamma _5[{\rm e}^{-\varepsilon \cdot \stackrel{\rightarrow }{\widetilde{D}}%
},\gamma \cdot \stackrel{\rightarrow }{D}]{\rm e}^{\varepsilon \cdot
\stackrel{\rightarrow }{\widetilde{D}}}-{\rm e}^{-\varepsilon \cdot
\stackrel{\leftarrow }{\widetilde{D}}}[\gamma \cdot \stackrel{\leftarrow }{D}%
,{\rm e}^{\varepsilon \cdot \stackrel{\leftarrow }{\widetilde{D}}}]\gamma
_5\alpha )\widetilde{\psi }_\varepsilon \right) ,
\end{eqnarray}
that means we have
\begin{equation}
\widetilde{{\cal A}}_5^a=\langle \widetilde{\psi }_\varepsilon ^{+}(\alpha
\gamma _5[{\rm e}^{-\varepsilon \cdot \stackrel{\rightarrow }{\widetilde{D}}%
},\gamma \cdot \stackrel{\rightarrow }{D}]{\rm e}^{\varepsilon \cdot
\stackrel{\rightarrow }{\widetilde{D}}}-{\rm e}^{-\varepsilon \cdot
\stackrel{\leftarrow }{\widetilde{D}}}[\gamma \cdot \stackrel{\leftarrow }{D}%
,{\rm e}^{\varepsilon \cdot \stackrel{\leftarrow }{\widetilde{D}}}]\gamma
_5\alpha )\widetilde{\psi }_\varepsilon \rangle .
\end{equation}
It is an easy exercise to prove, that this expression reproduces the formula
(\ref{chiral tilde}).


\section{Properties of the vector current}


In the previous section we have defined the covariant vector current by
means of the formula
\begin{equation}
j_\mu ^{a,{\rm reg}}=\psi _\varepsilon ^{+}T^a\gamma _\mu \psi _\varepsilon .
\label{ourj}
\end{equation}
From this formula it follows immediately that the regularized current
transforms covariantly under the gauge transformations. There exist also
other definitions of the covariant point-split currents, which differ from
this by the order of the parallel transporter and the gauge generator $T^a$.
E.g. in ref. \cite{modification} the authors use the prescription
\begin{equation}
j_{\mu ,(1)}^{a,{\rm reg}}=\psi ^{+}(x+\varepsilon )T^a\gamma _\mu \Omega
(x+\varepsilon ,x-\varepsilon )\psi (x-\varepsilon )  \label{j1}
\end{equation}
or
\begin{equation}
j_{\mu ,(2)}^{a,{\rm reg}}=\frac 12\psi ^{+}(x+\varepsilon )\{T^a,\Omega
(x+\varepsilon ,x-\varepsilon )\}\gamma _\mu \psi (x-\varepsilon ).
\label{j2}
\end{equation}
Let us now clarify the relations between these currents. We can write
\begin{equation}
j_{\mu ,(1)}^{a,{\rm reg}}=\psi _\varepsilon ^{+}{\rm e}^{-\varepsilon \cdot
\stackrel{\leftarrow }{D}}{\rm e}^{\varepsilon \cdot \stackrel{\leftarrow }{%
\partial }}T^a\gamma _\mu \Omega (x+\varepsilon ,x)\psi _\varepsilon ,
\end{equation}
or, using the identity
\begin{equation}
\Omega (x+\varepsilon ,x)={\rm e}^{-\varepsilon \cdot \stackrel{\leftarrow }{%
\partial }}{\rm e}^{\varepsilon \cdot \stackrel{\leftarrow }{D}},
\end{equation}
we have
\begin{eqnarray}
j_{\mu ,(1)}^{a,{\rm reg}} &=&\psi _\varepsilon ^{+}{\rm e}^{-\varepsilon
\cdot \stackrel{\leftarrow }{D}}{\rm e}^{\varepsilon \cdot \stackrel{%
\leftarrow }{\partial }}T^a\gamma _\mu {\rm e}^{-\varepsilon \cdot \stackrel{%
\leftarrow }{\partial }}{\rm e}^{\varepsilon \cdot \stackrel{\leftarrow }{D}%
}\psi _\varepsilon  \nonumber \\
\ &=&\psi _\varepsilon ^{+}{\rm e}^{-\varepsilon \cdot \stackrel{\leftarrow
}{D}}T^a\gamma _\mu {\rm e}^{\varepsilon \cdot \stackrel{\leftarrow }{D}%
}\psi _\varepsilon .
\end{eqnarray}
That means the current (\ref{ourj}) and (\ref{j1}) differ by the counterterm
\begin{eqnarray}
\Delta _{(1)}j_\mu ^a &=&\langle \psi _\varepsilon ^{+}({\rm e}%
^{-\varepsilon \cdot \stackrel{\leftarrow }{D}}T^a{\rm e}^{\varepsilon \cdot
\stackrel{\leftarrow }{D}}-T^a)\gamma _\mu \psi _\varepsilon \rangle
\nonumber \\
\ &=&-{\rm Tr}S^{{\rm cov}}\gamma _\mu ({\rm e}^{-\varepsilon \cdot
\stackrel{\leftarrow }{D}}T^a{\rm e}^{\varepsilon \cdot \stackrel{\leftarrow
}{D}}-T^a).
\end{eqnarray}
Because
\begin{eqnarray}
{\rm e}^{-\varepsilon \cdot \stackrel{\leftarrow }{D}}T^a{\rm e}%
^{\varepsilon \cdot \stackrel{\leftarrow }{D}}-T^a=\sum_{n=1}^\infty \frac
1{n!}[\varepsilon \cdot D,[\varepsilon \cdot D,[\cdots [\varepsilon \cdot
D,T^a]\ldots ]],
\end{eqnarray}
we have
\begin{eqnarray}
\Delta _{(1)}j_\mu ^a &=&\frac 1{4\pi ^2}{\rm Tr}_C\biggl(\left( \frac{\varepsilon
_\mu }{|\varepsilon |^4}-m^2\frac{\varepsilon _\mu }{|\varepsilon |^2}+{\cal %
O}(\ln |\varepsilon |)\right)
\nonumber \\
&&\times ([\varepsilon \cdot D,T^a]+\frac 12[\varepsilon \cdot
D,[\varepsilon \cdot D,T^a]]+\frac 16[\varepsilon \cdot D,[\varepsilon \cdot
D[\varepsilon \cdot D,T^a]]]+{\cal O}(\varepsilon ^4)) \biggr)
\nonumber \\
&=&{\cal O}(\varepsilon ,\varepsilon \ln |\varepsilon |),
\end{eqnarray}
in the same way we can easily prove
\begin{equation}
\Delta _{(2)}j_\mu ^a=j_{\mu ,(2)}^{a,{\rm reg}}-j_\mu ^{a,{\rm reg}}={\cal O%
}(\varepsilon ).
\end{equation}
That means that all the above introduced currents are equivalent in the
limit of the removed point-splitting (cf. \cite{modification}). But only one
of them, namely $j_\mu ^{a,{\rm reg}}(x)$, transforms covariantly according
to the naive prescription
\begin{equation}
j_\mu (x)\rightarrow U(x)j_\mu (x)U^{+}(x)
\end{equation}
already on the regularized level, the other currents have shifted the space
time argument of the transformation matrix $U$.

Let us investigate also another property of the general regularized current $%
j_\mu ^{a,{\rm reg}}(x).$ Provided the integrability conditions
\begin{equation}
I_{\mu \nu }^{ab}(x,y)[A]=\frac \delta {\delta A_\nu ^b(y)}\langle j_\mu ^{a,%
{\rm reg}}(x)\rangle _c-\frac \delta {\delta A_\mu ^a(x)}\langle j_\nu ^{b,%
{\rm reg}}(y)\rangle _c=0,  \label{integrability}
\end{equation}
are satisfied, the current can be expressed as a functional derivative of
the effective action functional $\Gamma _E[A]=\ln Z_E[A]$, i.e.
\begin{equation}
j_\mu ^{a,{\rm reg}}(x)\cdot Z_E[A]=\langle j_\mu ^{a,{\rm reg}}(x)\rangle
_c=-\frac \delta {\delta A_\mu ^a(x)}\Gamma _E[A]_c  \label{dGamma/dA}
\end{equation}
or, equivalently, it can be integrated to define the consistent effective
action consistent with the relation (\ref{dGamma/dA})
\begin{equation}
\Gamma _E[A]=-\int_0^1{\rm d}t\int {\rm d}^4xA_\mu ^a(x)j_\mu ^{a,{\rm reg}%
}(x)\cdot Z_E[tA]_c.  \label{effective action}
\end{equation}
$j_\mu ^{a,{\rm reg}}(x)\cdot Z_E[A]_c$ can be then interpreted as a
consistent generator of the correlators of the vector currents (i.e. which
obeys Bose symmetry) corresponding to the naive formula
\begin{equation}
\langle j_{\mu _1}^{a_1}(x_1)j_{\mu _2}^{a_2}(x_2)\ldots j_{\mu
_n}^{a_n}(x_n)\rangle _c
=(-1)^{n-1}\frac{\delta ^{n-1}}{\delta A_{\mu
_2}^{a_2}(x_2)\cdots \delta A_{\mu _n}^{a_n}(x_n)}j_{\mu_1} ^{a_1,{\rm reg}%
}(x_1)\cdot Z_E[A]_c.
\end{equation}
On the other hand, provided the right hand side of (\ref{integrability}) is
nonzero, the effective action defined by means of the integral (\ref
{effective action}) does not yield the original current, but the integrable
one (cf. also \cite{modification} and \cite{integrability})
\begin{equation}
J_\mu ^{a,{\rm reg}}(x)\cdot Z_E[A]=-\frac \delta {\delta A_\mu ^a(x)}\Gamma
_E[A]_c=j_\mu ^{a,{\rm reg}}(x)\cdot Z_E[A]-\int_0^1{\rm d}t\,t\int {\rm d}%
^4yI_{\mu \nu }^{ab}(x,y)[tA]A_\nu ^b(y)  \label{general integrable}
\end{equation}
the covariant derivative of which gives the consistent anomaly satisfying
the Wess-Zumino consistency conditions \cite{Wess-Zumino}.

Another characteristics of the general point-split current is the violation
of the covariance
\begin{equation}
C_\mu ^{ab}(x,y)[A]=D_\nu ^{ac}(y)\frac \delta {\delta A_\nu ^c(y)}j_\mu ^{b,%
{\rm reg}}(x)\cdot Z_E[A]_c-\delta ^{(4)}(x-y)f^{acb}j_\mu ^{c,{\rm reg}%
}(x)\cdot Z_E[A]_c.  \label{covariance}
\end{equation}
where $D_\nu ^{ac}(y)=\partial _\nu \delta ^{ac}+f^{dca}A_\nu ^d(y)$ is the
covariant derivative in the adjoint representation and $f^{dca}$ are the
totally antisymmetric structure constants of the gauge group. $C_\mu
^{ab}(x,y)[A]$ can be rewritten using (\ref{integrability}) to the form (cf.
ref. \cite{integrability})
\begin{eqnarray}
C_\mu ^{ab}(x,y)[A] &=&D_\nu ^{ac}(y)I_{\mu \nu }^{bc}(x,y)[A]+\frac \delta
{\delta A_\mu ^b(x)}D_\nu ^{ac}(y)j_\nu ^{c,{\rm reg}}(y)\cdot Z_E[A]_c
\nonumber \\
\ &=&-D_\nu ^{ac}(y)I_{\nu \mu }^{cb}(y,x)[A]+\frac \delta {\delta A_\mu
^b(x)}{\cal A}^a(y)[A]
\end{eqnarray}
where ${\cal A}^a(y)[A]=D_\nu ^{ac}(y)\langle j_\nu ^{c,{\rm reg}}(y)\rangle
_c$ is the anomaly. From this equation it is easily seen, that anomaly free
integrable current is necessary covariant and covariant and integrable
current must be anomaly free.

As an illustration, we prove in the following that for the covariant
point-splitting described in the previous sections the integrability
conditions (\ref{integrability}) are not violated (cf. also \cite
{modification}). Let us decompose the covariantly regularized current (\ref
{ourj}) into two parts, namely
\begin{equation}
j_\mu ^{a,{\rm reg}}=\widetilde{j}_\mu ^{a,{\rm reg}}-\Delta \widetilde{j}%
_\mu ^{a,{\rm reg}},
\end{equation}
where $\widetilde{j}_\mu ^{a,{\rm reg}}$ is the non-covariantly regularized
vector current (\ref{jnoncov}) described in the previous section and $\Delta
\widetilde{j}_\mu ^{a,{\rm reg}}$ is the corresponding counterterm and
choose appropriately the value of the regularization parameter $a$ in order
to simplify the calculations of $I_{\mu \nu }^{ab}(x,y)$ as much as
possible. The contribution $\Delta \widetilde{I}_{\mu \nu }^{ab}(x,y)$ of
the counterterm $\Delta \widetilde{j}_\mu ^{a,{\rm reg}}$ to the left hand
side of the integrability condition (\ref{integrability}) can be easily
obtained from the formula (\ref{jctnoncov}) as
\begin{eqnarray}
\Delta \widetilde{I}_{\mu \nu }^{ab}(x,y) &=&\frac \delta {\delta A_\nu
^b(y)}\langle \Delta \widetilde{j}_\mu ^{a,{\rm reg}}(x)\rangle _c-\frac
\delta {\delta A_\mu ^a(x)}\langle \Delta \widetilde{j}_\nu ^{b,{\rm reg}%
}(y)\rangle _c  \nonumber \\
\ &=&\frac{(1-a)(2-a)}{96\pi ^2}\delta ^{(4)}(x-y){\rm Tr}_C[T^a,T^b](\delta
_{\mu \nu }\partial \cdot A+\partial _\mu A_\nu +\partial _\nu A_\mu )
\nonumber \\
&&\ +{\cal O}(\varepsilon ,\varepsilon \ln |\varepsilon |).
\label{noncovariant integrability}
\end{eqnarray}
The contribution of the non-covariantly regularized vector current, $%
\widetilde{I}_{\mu \nu }^{ab}(x,y)$, can be rewritten in the form
\begin{eqnarray}
\widetilde{I}_{\mu \nu }^{ab}(x,y) &=&\frac \delta {\delta A_\nu
^b(y)}\langle \widetilde{j}_\mu ^{a,{\rm reg}}(x)\rangle _c-\frac \delta
{\delta A_\mu ^a(x)}\langle \widetilde{j}_\nu ^{b,{\rm reg}}(y)\rangle _c
\nonumber \\
\ &=&-\langle \widetilde{\psi }_\varepsilon ^{+}(x)T^a\gamma _\mu \widetilde{%
\psi }_\varepsilon (x)\psi ^{+}(y)T^b\gamma _\nu \psi (y)\rangle _c+\langle
\psi ^{+}(x)T^a\gamma _\mu \psi (x)\widetilde{\psi }_\varepsilon
^{+}(y)T^b\gamma _\nu \widetilde{\psi }_e(y)\rangle _c  \nonumber \\
&&\ \ \ \ +\langle \psi ^{+}(x)(\frac \delta {\delta A_\nu ^b(y)}{\rm e}%
^{\varepsilon \cdot \stackrel{\leftarrow }{\widetilde{D}}})T^a\gamma _\mu
\widetilde{\psi }_\varepsilon (x)\rangle _c+\langle \widetilde{\psi }%
_\varepsilon ^{+}(x)T^a\gamma _\mu (\frac \delta {\delta A_\nu ^b(y)}{\rm e}%
^{-\varepsilon \cdot \widetilde{D}})\psi (x)\rangle _c  \nonumber \\
&&\ \ \ \ -\langle \psi ^{+}(y)(\frac \delta {\delta A_\mu ^a(x)}{\rm e}%
^{\varepsilon \cdot \stackrel{\leftarrow }{\widetilde{D}}})T^b\gamma _\mu
\widetilde{\psi }_\varepsilon (y)\rangle _c-\langle \widetilde{\psi }%
_\varepsilon ^{+}(y)T^b\gamma _\mu (\frac \delta {\delta A_\mu ^a(x)}{\rm e}%
^{-\varepsilon \cdot \widetilde{D}})\psi (y)\rangle _c.  \nonumber \\
&&
\end{eqnarray}
Putting in this formula the value of the regularization parameter $a=0$, we
get much less complicated expression
\begin{eqnarray}
\widetilde{I}_{\mu \nu }^{ab}(x,y)|_{a=0} &=&-\langle \psi
^{+}(x+\varepsilon )T^a\gamma _\mu \psi (x-\varepsilon )\psi
^{+}(y)T^b\gamma _\nu \psi (y)\rangle _c  \nonumber \\
&&\ +\langle \psi ^{+}(x)T^a\gamma _\mu \psi (x)\psi ^{+}(y+\varepsilon
)T^b\gamma _\nu \psi (y-\varepsilon )\rangle _c  \nonumber \\
\ &=&{\rm Tr}T^a\gamma _\mu S(x-\varepsilon ,y)T^b\gamma _\nu
S(y,x+\varepsilon )  \nonumber \\
&&\ -{\rm Tr}T^a\gamma _\mu S(x,y+\varepsilon )T^b\gamma _\nu
S(y-\varepsilon ,x).
\end{eqnarray}
For $x\neq y$ the right hand side is easily seen to vanish in the limit $%
\varepsilon \rightarrow 0$, however for $x=y$ we get singular expression;
i.e. $\widetilde{I}_{\mu \nu }^{ab}(x,y)|_{a=0}$ as a function of $y$ is a
distribution concentrated at $y=x$. Let us therefore calculate the
expression
\begin{eqnarray}
\int {\rm d}^4yB_\nu ^b(y)\widetilde{I}_{\mu \nu }^{ab}(x,y)|_{a=0} &=&\int
{\rm d}^4y({\rm Tr}T^a\gamma _\mu S(x-\varepsilon ,y)B(y)\cdot \gamma
S(y,x+\varepsilon )  \nonumber \\
&&\ -{\rm Tr}T^a\gamma _\mu S(x,y+\varepsilon )B(y)\cdot \gamma
S(y-\varepsilon ,x)),
\end{eqnarray}
where $B_\nu ^b(y)$ are appropriately smooth test functions with compact
support. Inserting here the expansion of the propagator
\begin{eqnarray}
S(x,y) &=&\frac 1{4\pi ^2}\left(-2\frac{\gamma \cdot (x-y)}{|x-y|^4}+2\frac{%
\gamma \cdot (x-y)(x-y)\cdot A(\overline{x})}{|x-y|^4}-{\rm i}m\frac
1{|x-y|^2}
\right.
\nonumber \\
&&\ \ -\frac{\gamma \cdot (x-y)(x-y)\cdot A(\overline{x})A(\overline{x}%
)\cdot (x-y)}{|x-y|^4}+\frac 12m^2\frac{\gamma \cdot (x-y)}{|x-y|^2}
\nonumber \\
&&
\left.
\ \ +\frac 12\frac{(x-y)_\alpha F_{\alpha \beta }^{*}(\overline{x})\gamma
_\beta \gamma _5}{|x-y|^2}-{\rm i}m\frac{(x-y)\cdot A(\overline{x})}{|x-y|^2}%
+{\cal O}(\ln |\varepsilon |)\right),  \label{noncovS}
\end{eqnarray}
where $\overline{x}=(x+y)/2$, we find after the substitution $%
y=x+|\varepsilon |z$ , that the potentially most singular term stemming from
the product of the leading singularity in the expansion of $S(x,y)$ cancel
and the only terms, which survives the limit $\varepsilon \rightarrow 0$
comes from the product of the first two terms of (\ref{noncovS}), namely
\begin{eqnarray}
\int {\rm d}^4yB_\nu ^b(y)\widetilde{I}_{\mu \nu }^{ab}(x,y)|_{a=0}
&=&-\frac 1{4\pi ^4}{\rm Tr}_D(\gamma _\mu \gamma _\alpha \gamma _\nu \gamma
_\beta )\frac 1{|\varepsilon |}\int {\rm d}^4zB_\nu ^b(x+|\varepsilon |z)%
\frac{\Delta _{+\alpha }\Delta _{-\beta }}{|\Delta _{+}|^4|\Delta _{-}|^4}
\nonumber \\
&&\ \ \times ({\rm Tr}_C(T^aT^b(A_\gamma (\overline{x}_{+})-A_\gamma (%
\overline{x}_{-}))\Delta _{-\gamma })
\nonumber \\
&&\ \ +{\rm Tr}_C(T^bT^a(A_\gamma (\overline{x}_{+})-A_\gamma (\overline{x}%
_{-}))\Delta _{+\gamma }))+{\cal O}(\varepsilon \ln |\varepsilon |),
\end{eqnarray}
here we denote
\begin{eqnarray}
\Delta _{\pm } &=&\frac{x-y\mp \varepsilon }{|\varepsilon |}=\mp (z\pm n) \\
\overline{x}_{\pm } &=&\frac 12(x+y\pm \varepsilon )=x+\frac 12|\varepsilon
|(z\pm n)
\end{eqnarray}
and $n=\varepsilon /|\varepsilon |$ . We have therefore, after the
substitution $z\rightarrow -z$ in the second term
\begin{eqnarray}
\int {\rm d}^4yB_\nu ^b(y)\widetilde{I}_{\mu \nu }^{ab}(x,y)|_{a=0}
&=&-\frac 1{4\pi ^4}{\rm Tr}_D(\gamma _\mu \gamma _\alpha \gamma _\nu \gamma
_\beta )B_\nu ^b(x){\rm Tr}_C([T^a,T^b]\partial _\delta A_\gamma )n_\delta
J_{\alpha \beta \gamma }(n)  \nonumber \\
&&\ +{\cal O}(\varepsilon \ln |\varepsilon |),
\end{eqnarray}
where
\begin{eqnarray}
J_{\alpha \beta \gamma }(n) &=&\int {\rm d}^4z\frac{\Delta _{+\alpha }\Delta
_{-\beta }\Delta _{-\gamma }}{|\Delta _{+}|^4|\Delta _{-}|^4}  \nonumber \\
\ &=&\frac{\pi ^2}8(\delta _{\alpha \beta }n_\gamma -\delta _{\beta \gamma
}n_\alpha +\delta _{\alpha \gamma }n_\beta -2n_\alpha n_\beta n_\gamma ).
\end{eqnarray}
I.e. after taking the average over the directions of $n$ and some algebra we
get
\begin{eqnarray}
\widetilde{I}_{\mu \nu }^{ab}(x,y)|_{a=0} &=&\frac 1{48\pi ^2}\delta
^{(4)}(x-y){\rm Tr}_C([T^a,T^b](\delta _{\mu \nu }\partial \cdot A+\partial
_\mu A_\nu +\partial _\nu A_\mu ))
\nonumber \\
&&\ +{\cal O}(\varepsilon ,\varepsilon \ln |\varepsilon |)
\end{eqnarray}
and\footnote{%
We could also obtain this result in a more direct way and get independent
check of the calculation. Let us write
\begin{eqnarray*}
I_{\mu \nu }^{ab}(x,y)=I_{\mu \nu }^{ab}(x,y)^{(1)}+I_{\mu \nu
}^{ab}(x,y)^{(2)}
\end{eqnarray*}
where
\begin{eqnarray*}
I_{\mu \nu }^{ab}(x,y)^{(1)} &=&{\rm Tr}T^a\gamma _\mu \Omega
(x,x-\varepsilon )S(x-\varepsilon ,y)T^b\gamma _\nu S(y,x+\varepsilon
)\Omega (x+\varepsilon ,x) \\
&&\ -{\rm Tr}T^b\gamma _\nu \Omega (y,y-\varepsilon )S(y-\varepsilon
,x)T^a\gamma _\mu S(x,y+\varepsilon )\Omega (y+\varepsilon ,y)
\end{eqnarray*}
and
\begin{eqnarray*}
I_{\mu \nu }^{ab}(x,y)^{(2)} &=&{\rm Tr}S^{{\rm cov}}(x,\varepsilon )\gamma
_\mu (\Delta _\nu ^b(x,y,-\varepsilon )T^a-T^a\Delta _\nu ^b(x,y,\varepsilon
)) \\
&&\ -{\rm Tr}S^{{\rm cov}}(y,\varepsilon )\gamma _\nu (\Delta _\mu
^a(y,x,-\varepsilon )T^b-T^b\Delta _\mu ^a(y,x,\varepsilon ))
\end{eqnarray*}
where
\begin{eqnarray*}
\Delta _\nu ^b(x,y,\varepsilon )
&=&
\left(\frac \delta {\delta A_\nu ^b(y)}
{\rm e}^{-\varepsilon \cdot D}\right) {\rm e}^{\varepsilon \cdot D}
=-\int_0^1{\rm d}t{\rm e}^{-t\varepsilon \cdot D}
\varepsilon _\nu \delta ^{(4)}(x-y)T^b
{\rm e}^{t\varepsilon \cdot D}
\\
\ &=&\sum_{n=0}^\infty \frac{(-1)^{n+1}}{(n+1)!}
[\varepsilon \cdot D,[\varepsilon \cdot D,[\cdots
[\varepsilon \cdot D,\varepsilon _\nu \delta^{(4)}(x-y)T^b]\ldots ]].
\end{eqnarray*}
Using the same method as in the main text, we can easily prove that $I_{\mu
\nu }^{ab}(x,y)^{(1)}=0.$ We have therefore, dropping terms which vanish
after symmetrization over the directions of the four-vector $\varepsilon $,
\begin{eqnarray*}
I_{\mu \nu }^{ab}(x,y)^{(2)}
&=&-\frac 1{4\pi ^2}\left(
\frac{\varepsilon_\mu \varepsilon _\nu }{|\varepsilon |^4}
-m^2\frac{\varepsilon _\mu \varepsilon _\nu }{|\varepsilon |^2}
+{\cal O}(\ln |\varepsilon |)\right)
\\
&&\ \ \times {\rm Tr}_C\left(
\frac 23\varepsilon \cdot \partial \delta^{(4)}(x-y)
[\varepsilon \cdot A(x),T^b]T^a
\right.
\\
&&\ \ +\frac 13\delta ^{(4)}(x-y)[\varepsilon \cdot D,[\varepsilon \cdot
A(x),T^b]]T^a
\\
&&
\left.
\ \ +\frac 13(\varepsilon \cdot \partial )^2\delta ^{(4)}(x-y)T^bT^a
+{\cal O}(\varepsilon ^3)\right)-((x,a)\leftrightarrow (y,b))
\\
\ &=&{\cal O}(\varepsilon ,\varepsilon \ln |\varepsilon |),
\end{eqnarray*}
i.e.
\[
I_{\mu \nu }^{ab}(x,y)=
{\cal O}(\varepsilon ,\varepsilon \ln |\varepsilon|).
\]
}
\begin{eqnarray}
I_{\mu \nu }^{ab}(x,y)=\widetilde{I}_{\mu \nu }^{ab}(x,y)|_{a=0}-\Delta
\widetilde{I}_{\mu \nu }^{ab}(x,y)|_{a=0}={\cal O}(\varepsilon ,\varepsilon
\ln |\varepsilon |).  \nonumber
\end{eqnarray}
That means the covariantly regularized vector current is integrable and,
because its divergence is zero, the effective action (\ref{effective action}%
) is gauge invariant.

As it is seen from formula (\ref{noncovariant integrability}), for the
non-covariant point-splitting
\begin{eqnarray}
\widetilde{I}_{\mu \nu }^{ab}(x,y) &=&\Delta \widetilde{I}_{\mu \nu
}^{ab}(x,y)=\frac{(1-a)(2-a)}{96\pi ^2}\delta ^{(4)}(x-y){\rm Tr}%
_C[T^a,T^b](\delta _{\mu \nu }\partial \cdot A+\partial _\mu A_\nu +\partial
_\nu A_\mu )  \nonumber \\
&&+{\cal O}(\varepsilon ,\varepsilon \ln |\varepsilon |)
\end{eqnarray}
and therefore we can recover integrability also for $a=2$. We can also
enlarge the class of regularizations and construct one parametric set of
integrable non-covariant vector currents within such an enlarged
point-splitting scheme in the following way. Let us define the regularized
current to be
\begin{equation}
j_\mu ^{a,{\rm reg}}(x;a,b)=\frac 12(\widetilde{\psi }_\varepsilon
^{+}(x,a)T^a\gamma _\mu \widetilde{\psi }_\varepsilon (x,b)+\widetilde{\psi }%
_\varepsilon ^{+}(x,b)T^a\gamma _\mu \widetilde{\psi }_\varepsilon (x,a))
\end{equation}
where

\begin{eqnarray}
\widetilde{\psi }_\varepsilon (x,a) &=&{\rm e}^{-\varepsilon \cdot D(a)}\psi
(x)  \nonumber \\
\widetilde{\psi }_\varepsilon ^{+}(x,a) &=&\psi (x){\rm e}^{\varepsilon
\cdot \stackrel{\leftarrow }{D}(a)}
\end{eqnarray}
and $D(a)=\partial +aA.$ This broad class of regularizations naturally
incorporates the previously introduced currents for special choice of the
parameters $a$ and $b$, namely $\widetilde{j}_\mu ^{a,{\rm reg}}(x)=j_\mu
^{a,{\rm reg}}(x;a,a)$ and $j_\mu ^{a,{\rm reg}}(x)=j_\mu ^{a,{\rm reg}%
}(x;1,1)$. It is not difficult to calculate the counterterm
\begin{equation}
\Delta j_\mu ^{a,{\rm reg}}(x;a,b)=j_\mu ^{a,{\rm reg}}(x;a,b)-j_\mu ^{a,%
{\rm reg}}(x),
\end{equation}
using the methods described in the previous section. We have
\begin{equation}
\Delta j_\mu ^{a,{\rm reg}}(x;a,b)=-\frac 12{\rm Tr}S^{{\rm cov}}\gamma _\mu
(\Omega _{+}(a)T^a\Omega _{-}(b)+\Omega _{+}(b)T^a\Omega _{-}(a)-2T^a)
\end{equation}
where we explicitly picked up the value of the parameters in the functions $%
\Omega _{\pm }$. Using the formulae of Appendix B we get
\begin{eqnarray}
\frac 12(\Omega _{-}(a)\Omega _{+}(b)+\Omega _{-}(b)\Omega _{+}(a)) &=&
1+2\left(1-\frac{a+b}2\right)\varepsilon \cdot A
+2\left(1-\frac{a+b}2\right)^2(\varepsilon \cdot A)^2
\nonumber \\
&&\ \ +\frac 43\left(1-\frac{a+b}2\right)^3(\varepsilon \cdot A)^3
+\frac 13\left(1-\frac{a+b}2\right)(\varepsilon \cdot \partial )^2
\varepsilon \cdot A
\nonumber \\
&&\ \ +\frac 1{12}((1-a)(a-3b+4)+(1-b)(b-3a+4))  \nonumber \\
&&\ \times [\varepsilon \cdot A,\varepsilon \cdot \partial A\cdot
\varepsilon ]+{\cal O}(\varepsilon ^4)
\end{eqnarray}
and therefore
\begin{eqnarray}
\Delta j_\mu ^{a,{\rm reg}}(x;a,b)
&=&\frac 1{8\pi ^2}{\rm Tr}_CT^a\left(\left(1-\frac{a+b}2\right)
\left(\frac 1{|\varepsilon |^2}-m^2\right)A_\mu
\right.
\nonumber \\
&&\ \ \ \ \ \ \ +\frac 19\left(1-\frac{a+b}2\right)^3(A_\mu A\cdot A+A\cdot AA_\mu
+A_\nu A_\mu A_\nu )
\nonumber \\
&&\ \ \ \ \ \ \ +\frac 1{144}((1-a)(a-3b+4)+(1-b)(b-3a+4))
\nonumber \\
&&\ \times ([A_\mu ,\partial \cdot A]+[A_\nu ,\partial _\mu A_\nu ]+[A_\nu
,\partial _\nu A_\mu ])
\nonumber \\
&&
\left.
\ \ \ \ \ \ \ +\frac 1{36}\left(1-\frac{a+b}2\right)
(2\partial _\mu \partial \cdot A+\partial ^2A_\mu )\right)
+{\cal O}(\varepsilon ).
\end{eqnarray}
This generally nonintegrable counterterm is ``minimal'' for $a+b=2$ , in
this case
\begin{equation}
\Delta j_\mu ^{a,{\rm reg}}(x;a,2-a)=-\frac{(1-a)^2}{144\pi ^2}{\rm Tr}%
_CT^a([A_\mu ,\partial \cdot A]+[A_\nu ,\partial _\mu A_\nu ]+[A_\nu
,\partial _\nu A_\mu ]),  \label{minimal counterterm}
\end{equation}
This minimal choice is exceptional also for another reason. Let us note,
that provided we regularize the axial current in the analogous way, i.e. if
we put
\begin{equation}
j_{\mu 5}^{a,{\rm reg}}(x;a,b)=\frac 12(\widetilde{\psi }_\varepsilon
^{+}(x,a)T^a\gamma _\mu \gamma _5\widetilde{\psi }_\varepsilon (x,b)+%
\widetilde{\psi }_\varepsilon ^{+}(x,b)T^a\gamma _\mu \gamma _5\widetilde{%
\psi }_\varepsilon (x,a)),
\end{equation}
we get the following non-covariant counterterm, which modifies the
covariantly regularized current $j_{\mu 5}^{a,{\rm reg}}(x)=j_{\mu 5}^{a,%
{\rm reg}}(x;1,1)$
\begin{eqnarray}
\Delta j_{\mu 5}^{a,{\rm reg}}(x;a,b) &=&j_{\mu 5}^{a,{\rm reg}%
}(x;a,b)-j_{\mu 5}^{a,{\rm reg}}  \nonumber \\
\ &=&-\frac 12{\rm Tr}S^{{\rm cov}}\gamma _\mu \gamma _5(\Omega
_{+}(a)T^a\Omega _{-}(b)+\Omega _{+}(b)T^a\Omega _{-}(a)-2T^a)  \nonumber \\
\ &=&\frac 1{16\pi ^2}\left(1-\frac{a+b}2\right){\rm Tr}_CT^a\{A_\nu ,F_{\nu \mu
}^{*}\}+{\cal O}(\varepsilon ).
\end{eqnarray}
This counterterm vanishes and the gauge covariance is recovered in the limit
of the removed cut-off again for $a$ and $b$ satisfying $a+b=2$.

The non-covariant current $j_\mu ^{a,{\rm reg}}(x;a,b)$ can be made
integrable for $a$ and $b$ for which
\begin{equation}
(1-a)(a-3b+4)+(1-b)(b-3a+4)=0;
\end{equation}
especially for the ``minimal'' choice $a=1$ ( however then $b=2-a=1$ and the
gauge covariance is recovered ). The gauge invariance of the effective
action $\widetilde{\Gamma }[A]\,$integrated from such a current is generally
lost, its variation under gauge transformation is determined by the
non-covariant anomaly,
\begin{eqnarray}
\widetilde{\Gamma }[A-[D,\alpha ]]
&=&\frac 1{8\pi ^2}\left(1-\frac{a+b}2\right)\int
{\rm d}^4x{\rm Tr}_C\alpha (x)[D_\mu ,((\frac 1{|\varepsilon |^2}-m^2)A_\mu
\nonumber \\
&&\ \ \ \ \ \ \ \ +\frac 19\left(1-\frac{a+b}2\right)^2(A_\mu A\cdot A+A\cdot AA_\mu
+A_\nu A_\mu A_\nu )  \nonumber \\
&&\ \ \ \ \ \ \ \ +\frac 1{36}(2\partial _\mu \partial \cdot A+\partial
^2A_\mu ))]+{\cal O}(\varepsilon ).
\end{eqnarray}

\vskip1cm {\large {\bf Acknowledgment }} \vskip1cm

The authors would like to thank J. Ho\v{r}ej\v{s}\'{\i} for valuable
discussions, careful reading of the manuscript and useful remarks related to
it.

This work was partially supported by the grants GAUK-166/95 and
GA\v{C}R-0506/98 .

\vskip1cm {\large {\bf Appendix A }} \nopagebreak\vskip1cm

In this appendix we give some technical details of the previous calculation.
We present here the explicit form of the functions $a_0(x,x-\varepsilon
)a_1(x-\varepsilon ,x+\varepsilon )a_0(x+\varepsilon ,x)$, $G_\mu
(x,\varepsilon )$ and $H_\mu (x,\varepsilon )$ used in the main text.

Let us first derive the formula (\ref{a0a1a0}). As we have seen in Section
\ref{asymptotic}, for $a_1(x,y)$ we have the following expression
\begin{equation}
a_1(x,y)=-\int_0^1{\rm d}ta_0(x,\widetilde{x}_t)\Delta _{\widetilde{x}_t}a_0(%
\widetilde{x}_t,y),
\end{equation}
where
\begin{equation}
\widetilde{x}_t=y+t(x-y)
\end{equation}
and
\begin{equation}
\Delta =-D^2-\frac{{\rm i}}2\sigma _{\mu \nu }F_{\mu \nu }.
\end{equation}
I.e. we have
\begin{eqnarray}
a_0(x,x-\varepsilon )a_1(x-\varepsilon ,x+\varepsilon )a_0(x+\varepsilon ,x)
&=&-\int_0^1{\rm d}ta_0(x,x_t)\Delta _{x_t}a_0(x_t,x+\varepsilon
)a_0(x+\varepsilon ,x)  \nonumber \\
\ &=&\frac{{\rm i}}2\int_0^1{\rm d}ta_0(x,x_t)\sigma _{\mu \nu }F_{\mu \nu
}(x_t)a_0(x_t,x)  \nonumber \\
&&\ +\int_0^1{\rm d}ta_0(x,x_t)(D_{x_t}^2a_0(x_t,x+\varepsilon
))a_0(x+\varepsilon ,x)  \nonumber \\
&&
\end{eqnarray}
where now
\begin{equation}
x_t=x+\varepsilon (1-2t).
\end{equation}
Let us remind that
\begin{equation}
a_0(x,y)=T\exp \left( -\int_0^1{\rm d}t(x-y)\cdot A(y+t(x-y))\right) ,
\end{equation}
i.e.
\begin{equation}
a_0(x,x_t)=T\exp \left( -\int_0^1{\rm d}\tau (x-x_t)\cdot A(x_t+\tau
(x-x_t))\right)
\end{equation}
and
\begin{equation}
a_0(x_t,x)=T\exp \left( \int_0^1{\rm d}\tau (x-x_t)\cdot A(x+\tau
(x_t-x))\right) .
\end{equation}
Expanding this to the first order in $\varepsilon $ we get
\begin{equation}
a_0(x,x_t)=1-(2t-1)\varepsilon \cdot A(x)+{\cal O}(\varepsilon ^2)
\end{equation}
and
\begin{equation}
a_0(x_t,x)=1+(2t-1)\varepsilon \cdot A(x)+{\cal O}(\varepsilon ^2).
\end{equation}
We have therefore immediately
\begin{eqnarray}
&&\frac{{\rm i}}2\int_0^1{\rm d}ta_0(x,x_t)\sigma _{\mu \nu }F_{\mu \nu
}(x_t)a_0(x_t,x)  \nonumber \\
&=&\frac{{\rm i}}2\int_0^1{\rm d}ta_0(x,x_t)\sigma _{\mu \nu }(F_{\mu \nu
}(x)+(1-2t)\varepsilon \cdot \partial F_{\mu \nu }(x))a_0(x_t,x)+{\cal O}%
(\varepsilon ^2)  \nonumber \\
&=&\frac{{\rm i}}2\sigma _{\mu \nu }F_{\mu \nu }(x)+\frac{{\rm i}}2\int_0^1%
{\rm d}t(1-2t)\varepsilon \cdot (\partial F_{\mu \nu }(x)+[A(x),F_{\mu \nu
}(x)])\sigma _{\mu \nu }+{\cal O}(\varepsilon ^2)  \nonumber \\
&=&\frac{{\rm i}}2\sigma _{\mu \nu }F_{\mu \nu }(x)+{\cal O}(\varepsilon ^2).
\end{eqnarray}
Let us calculate also two other quantities, namely $D_{x,\mu }a_0(x,y)$ and $%
D_x^2a_0(x,y).$ From the equation

\begin{equation}
(x-y)\cdot D_xa_0(x,y)=0
\end{equation}
we get
\begin{eqnarray}
&&D_{x,\mu }a_0(x,y)+(x-y)_\alpha D_{x,\mu }D_{x,\alpha }a_0(x,y)  \nonumber
\\
&=&(1+(x-y)\cdot D_x)D_{x,\mu }a_0(x,y)-(x-y)_\alpha F_{\alpha \mu
}a_0(x,y)=0
\end{eqnarray}
and, using the left hand side of the previous equation and the identity $%
[D^2,D_\alpha ]=[D_\mu ,F_{\mu \alpha }]+2F_{\mu \alpha }D_\mu $, we have
\begin{eqnarray}
&&2D_x^2a_0(x,y)+(x-y)_\alpha D_x^2D_{x,\alpha }a_0(x,y)  \nonumber \\
&=&(2+(x-y)\cdot D_x)D_x^2a_0(x,y)+(x-y)_\alpha ([D_\mu ,F_{\mu \alpha
}]+2F_{\mu \alpha }D_\mu )a_0(x,y)=0.
\end{eqnarray}
From this equations we get for $tD_{\widetilde{x}_t,\mu }a_0(\widetilde{x}%
_t,y)$ and $t^2D_{\widetilde{x}_t}^2a_0(\widetilde{x}_t,y)$ the following
ordinary differential equations
\begin{eqnarray}
\frac{{\rm d}}{{\rm d}t}tD_{\widetilde{x}_t,\mu }a_0(\widetilde{x}_t,y)
&=&-(x-y)\cdot A(\widetilde{x}_t)tD_{\widetilde{x}_t,\mu }a_0(\widetilde{x}%
_t,y)  \nonumber \\
&&+t(x-y)_\alpha F_{\alpha \mu }(\widetilde{x}_t)a_0(\widetilde{x}_t,y)
\end{eqnarray}
and
\begin{eqnarray}
\frac{{\rm d}}{{\rm d}t}t^2D_{\widetilde{x}_t}^2a_0(\widetilde{x}_t,y)
&=&-(x-y)\cdot A(\widetilde{x}_t)t^2D_{\widetilde{x}_t}^2a_0(\widetilde{x}%
_t,y)  \nonumber \\
&&-t^2(x-y)_\alpha ([D_{\widetilde{x}_t,\mu },F_{\mu \alpha }(\widetilde{x}%
_t)]+2F_{\mu \alpha }(\widetilde{x}_t)D_{\widetilde{x}_t,\mu })a_0(%
\widetilde{x}_t,y)
\end{eqnarray}
with the solutions
\begin{equation}
tD_{\widetilde{x}_t,\mu }a_0(\widetilde{x}_t,y)=(x-y)_\alpha \int_0^t{\rm d}%
\tau \tau a_0(\widetilde{x}_t,\widetilde{x}_\tau )F_{\alpha \mu }(\widetilde{%
x}_\tau )a_0(\widetilde{x}_\tau ,y)  \label{D}
\end{equation}
and
\begin{eqnarray}
t^2D_{\widetilde{x}_t}^2a_0(\widetilde{x}_t,y) &=&-(x-y)_\alpha \int_0^t{\rm %
d}\tau \tau ^2a_0(\widetilde{x}_t,\widetilde{x}_\tau )[D_{\widetilde{x}_\tau
,\mu },F_{\mu \alpha }(\widetilde{x}_\tau )]a_0(\widetilde{x}_\tau ,y)
\nonumber \\
&&-2(x-y)_\alpha \int_0^t{\rm d}\tau \tau a_0(\widetilde{x}_t,\widetilde{x}%
_\tau )F_{\mu \alpha }(\widetilde{x}_\tau )(\tau D_{\widetilde{x}_\tau ,\mu
}a_0(\widetilde{x}_\tau ,y)).  \label{DD}
\end{eqnarray}
Now, setting $x\rightarrow x-\varepsilon $, $y\rightarrow x+\varepsilon ,$
we can easily expand in powers of $\varepsilon .$ From (\ref{D}) we see that
up to the order ${\cal O}(\varepsilon )$ only the first term on the right
hand side of (\ref{DD}) contributes, i.e.
\begin{eqnarray}
t^2D_{x_t}^2a_0(x_t,y) &=&-2\varepsilon _\alpha \int_0^t{\rm d}\tau \tau
^2[D_\mu ,F_{\mu \alpha }(x)]+{\cal O}(\varepsilon ^2)  \nonumber \\
&=&-2\varepsilon _\alpha \frac{t^3}3[D_\mu ,F_{\mu \alpha }(x)]+{\cal O}%
(\varepsilon ^2).
\end{eqnarray}
As a result we have
\begin{eqnarray}
\int_0^1{\rm d}ta_0(x,x_t)(D_{x_t}^2a_0(x_t,x+\varepsilon
))a_0(x+\varepsilon ,x) &=&-2\varepsilon _\alpha \int_0^1{\rm d}t\frac
t3[D_\mu ,F_{\mu \alpha }(x)]+{\cal O}(\varepsilon ^2)  \nonumber \\
&=&-\frac 13\varepsilon _\alpha [D_\mu ,F_{\mu \alpha }(x)]+{\cal O}%
(\varepsilon ^2).
\end{eqnarray}
Finally we get
\begin{equation}
a_0(x,x-\varepsilon )a_1(x-\varepsilon ,x+\varepsilon )a_0(x+\varepsilon ,x)=%
\frac{{\rm i}}2\sigma _{\mu \nu }F_{\mu \nu }(x)+\frac 13\varepsilon _\alpha
[D_\mu ,F_{\mu \alpha }(x)]+{\cal O}(\varepsilon ^2).
\end{equation}

Now we present the derivation of the right hand side of the formula (\ref
{G-H formula}). Let us evaluate the following expression

\begin{eqnarray}
\partial _\mu ^\varepsilon ({\rm e}^{-\varepsilon \cdot \stackrel{%
\rightarrow }{D_x}}G(x,y;m^2){\rm e}^{\varepsilon \cdot \stackrel{\leftarrow
}{D}_y}) &=&-\int_0^1{\rm d}t{\rm e}^{-t\varepsilon \cdot \stackrel{%
\rightarrow }{D_x}}\stackrel{\rightarrow }{D}_{\mu ,x}{\rm e}^{t\varepsilon
\cdot \stackrel{\rightarrow }{D_x}}{\rm e}^{-\varepsilon \cdot \stackrel{%
\rightarrow }{D_x}}G(x,y;m^2){\rm e}^{\varepsilon \cdot \stackrel{\leftarrow
}{D}_y}  \nonumber \\
&&+\int_0^1{\rm d}t{\rm e}^{-\varepsilon \cdot \stackrel{\rightarrow }{D_x}%
}G(x,y;m^2){\rm e}^{\varepsilon \cdot \stackrel{\leftarrow }{D}_y}{\rm e}%
^{-t\varepsilon \cdot \stackrel{\leftarrow }{D}_y}\stackrel{\leftarrow }{D}%
_{\mu ,y}{\rm e}^{t\varepsilon \cdot \stackrel{\leftarrow }{D}_y}  \nonumber
\\
&=&-{\rm e}^{-\varepsilon \cdot \stackrel{\rightarrow }{D_x}}\stackrel{%
\rightarrow }{D}_{\mu ,x}G(x,y;m^2){\rm e}^{\varepsilon \cdot \stackrel{%
\leftarrow }{D}_y}+{\rm e}^{-\varepsilon \cdot \stackrel{\rightarrow }{D_x}%
}G(x,y;m^2)\stackrel{\leftarrow }{D}_{\mu ,y}{\rm e}^{\varepsilon \cdot
\stackrel{\leftarrow }{D}_y}  \nonumber \\
&&-\int_0^1{\rm d}tt{\rm e}^{-t\varepsilon \cdot \stackrel{\rightarrow }{D_x}%
}[\varepsilon \cdot \stackrel{\rightarrow }{D}_x,\stackrel{\rightarrow }{D}%
_{\mu ,x}]{\rm e}^{t\varepsilon \cdot \stackrel{\rightarrow }{D_x}}{\rm e}%
^{-\varepsilon \cdot \stackrel{\rightarrow }{D_x}}G(x,y;m^2){\rm e}%
^{\varepsilon \cdot \stackrel{\leftarrow }{D}_y}  \nonumber \\
&&+\int_0^1{\rm d}tt{\rm e}^{-\varepsilon \cdot \stackrel{\rightarrow }{D_x}%
}G(x,y;m^2){\rm e}^{\varepsilon \cdot \stackrel{\leftarrow }{D}_y}{\rm e}%
^{-t\varepsilon \cdot \stackrel{\leftarrow }{D}_y}[\varepsilon \cdot
\stackrel{\leftarrow }{D}_y,\stackrel{\leftarrow }{D}_{\mu ,y}]{\rm e}%
^{t\varepsilon \cdot \stackrel{\leftarrow }{D}_y},  \nonumber \\
&&
\end{eqnarray}
where we have used integration by parts to get the last lines. Introducing
the function
\begin{eqnarray}
H_\mu (x,\varepsilon ) &=&-\int_0^1{\rm d}tt{\rm e}^{-t\varepsilon \cdot
\stackrel{\rightarrow }{D_x}}[\varepsilon \cdot \stackrel{\rightarrow }{D}_x,%
\stackrel{\rightarrow }{D}_{\mu ,x}] {\rm e}^{t\varepsilon \cdot \stackrel{%
\rightarrow }{D_x}}
\nonumber \\
\ &=&-\int_0^1{\rm d}tt{\rm e}^{-t\varepsilon \cdot \stackrel{\rightarrow }{%
D_x}}\varepsilon _\nu F_{\nu \mu }(x){\rm e}^{t\varepsilon \cdot \stackrel{%
\rightarrow }{D_x}}  \nonumber \\
\ &=&\sum_{n=0}^\infty \frac{(-1)^{n+1}}{n!(n+2)}[\varepsilon \cdot
D,[\varepsilon \cdot D,[\ldots ,[\varepsilon \cdot D,\varepsilon _\nu F_{\nu
\mu }(x)]\ldots ]]].  \label{H function}
\end{eqnarray}
we can rewrite it in the form
\begin{eqnarray}
\partial _\mu ^\varepsilon ({\rm e}^{-\varepsilon \cdot \stackrel{%
\rightarrow }{D_x}}G(x,y;m^2){\rm e}^{\varepsilon \cdot \stackrel{\leftarrow
}{D}_y}) &=&-{\rm e}^{-\varepsilon \cdot \stackrel{\rightarrow }{D_x}}%
\stackrel{\rightarrow }{D}_{\mu ,x}G(x,y;m^2){\rm e}^{\varepsilon \cdot
\stackrel{\leftarrow }{D}_y}+{\rm e}^{-\varepsilon \cdot \stackrel{%
\rightarrow }{D_x}}G(x,y;m^2)\stackrel{\leftarrow }{D}_{\mu ,y}{\rm e}%
^{\varepsilon \cdot \stackrel{\leftarrow }{D}_y}  \nonumber \\
&&+H_\mu (x,\varepsilon ){\rm e}^{-\varepsilon \cdot \stackrel{\rightarrow }{%
D_x}}G(x,y;m^2){\rm e}^{\varepsilon \cdot \stackrel{\leftarrow }{D}_y}
\nonumber \\
&&+{\rm e}^{-\varepsilon \cdot \stackrel{\rightarrow }{D_x}}G(x,y;m^2){\rm e}%
^{\varepsilon \cdot \stackrel{\leftarrow }{D}_y}H_\mu (x,-\varepsilon ).
\end{eqnarray}
Analogously we have

\begin{eqnarray}
\lbrack D_\mu ,G^{{\rm cov}}(x,\varepsilon )] &=&\partial _\mu G^{{\rm cov}%
}(x,\varepsilon )+A_\mu(x)G^{{\rm cov}}(x,\varepsilon )-G^{{\rm cov}%
}(x,\varepsilon )A_\mu(x)
\nonumber \\
&=&\partial _\mu ^x{\rm e}^{-\varepsilon \cdot \stackrel{\rightarrow }{D_x}%
}G(x,y;m^2){\rm e}^{\varepsilon \cdot \stackrel{\leftarrow }{D}_y}+{\rm e}%
^{-\varepsilon \cdot \stackrel{\rightarrow }{D_x}}G(x,y;m^2){\rm e}%
^{\varepsilon \cdot \stackrel{\leftarrow }{D}_y}\stackrel{\leftarrow }{%
\partial }_\mu ^y  \nonumber \\
&&+A_\mu(x){\rm e}^{-\varepsilon \cdot \stackrel{\rightarrow }{D_x}}G(x,y;m^2)%
{\rm e}^{\varepsilon \cdot \stackrel{\leftarrow }{D}_y}-{\rm e}%
^{-\varepsilon \cdot \stackrel{\rightarrow }{D_x}}G(x,y;m^2){\rm e}%
^{\varepsilon \cdot \stackrel{\leftarrow }{D}_y}A_\mu(y)|_{x=y}
\nonumber \\
&=&D_{x,\mu }{\rm e}^{-\varepsilon \cdot \stackrel{\rightarrow }{D_x}%
}G(x,y;m^2){\rm e}^{\varepsilon \cdot \stackrel{\leftarrow }{D}_y}+{\rm e}%
^{-\varepsilon \cdot \stackrel{\rightarrow }{D_x}}G(x,y;m^2){\rm e}%
^{\varepsilon \cdot \stackrel{\leftarrow }{D}_y}\stackrel{\leftarrow }{D}%
_{y,\mu }|_{x=y}  \nonumber \\
&=&{\rm e}^{-\varepsilon \cdot \stackrel{\rightarrow }{D_x}}\stackrel{%
\rightarrow }{D}_{\mu ,x}G(x,y;m^2){\rm e}^{\varepsilon \cdot \stackrel{%
\leftarrow }{D}_y}+{\rm e}^{-\varepsilon \cdot \stackrel{\rightarrow }{D_x}%
}G(x,y;m^2)\stackrel{\leftarrow }{D}_{\mu ,y}{\rm e}^{\varepsilon \cdot
\stackrel{\leftarrow }{D}_y}|_{x=y}  \nonumber \\
&&+[\stackrel{\rightarrow }{D}_{\mu ,x},{\rm e}^{-\varepsilon \cdot
\stackrel{\rightarrow }{D_x}}]{\rm e}^{\varepsilon \cdot \stackrel{%
\rightarrow }{D_x}}G^{{\rm cov}}(x,\varepsilon )+G^{{\rm cov}}(x,\varepsilon
){\rm e}^{-\varepsilon \cdot \stackrel{\leftarrow }{D}_x}[{\rm e}%
^{\varepsilon \cdot \stackrel{\leftarrow }{D}_x},\stackrel{\leftarrow }{D}%
_{\mu ,x}].
\end{eqnarray}
and introducing the function
\begin{eqnarray}
G_\mu (x,\varepsilon ) &=&-[\stackrel{\rightarrow }{D}_{\mu ,x},{\rm e}%
^{-\varepsilon \cdot \stackrel{\rightarrow }{D_x}}]{\rm e}^{\varepsilon
\cdot \stackrel{\rightarrow }{D_x}}=\int_0^1{\rm d}t{\rm e}^{-t\varepsilon
\cdot \stackrel{\rightarrow }{D_x}}F_{\mu \nu }\varepsilon _\nu {\rm e}%
^{t\varepsilon \cdot \stackrel{\rightarrow }{D_x}}  \nonumber \\
\ &=&\sum_{n=0}^\infty \frac{(-1)^n}{(n+1)!}[\varepsilon \cdot
D,[\varepsilon \cdot D,[\ldots ,[\varepsilon \cdot D,\varepsilon _\nu F_{\mu
\nu }(x)]\ldots ]]]  \label{G function}
\end{eqnarray}
we have
\begin{eqnarray}
\lbrack D_\mu ,G^{{\rm cov}}(x,\varepsilon )] &=&{\rm e}^{-\varepsilon \cdot
\stackrel{\rightarrow }{D_x}}\stackrel{\rightarrow }{D}_{\mu ,x}G(x,y;m^2)%
{\rm e}^{\varepsilon \cdot \stackrel{\leftarrow }{D}_y}+{\rm e}%
^{-\varepsilon \cdot \stackrel{\rightarrow }{D_x}}G(x,y;m^2)\stackrel{%
\leftarrow }{D}_{\mu ,y}{\rm e}^{\varepsilon \cdot \stackrel{\leftarrow }{D}%
_y}|_{x=y}  \nonumber \\
&&\ \ \ -G_\mu (x,\varepsilon )G^{{\rm cov}}(x,\varepsilon )+G^{{\rm cov}%
}(x,\varepsilon )G_\mu (x,-\varepsilon ).
\end{eqnarray}
As a result we have
\begin{eqnarray}
{\rm e}^{-\varepsilon \cdot \stackrel{\rightarrow }{D_x}}\stackrel{%
\rightarrow }{D}_{\mu ,x}G(x,y;m^2){\rm e}^{\varepsilon \cdot \stackrel{%
\leftarrow }{D}_y}|_{x=y} &=&\frac 12([D_\mu ,G^{{\rm cov}}(x,\varepsilon
)]-\partial ^\varepsilon G^{{\rm cov}}(x,\varepsilon )  \nonumber \\
&&\ \ \ \ +G_\mu (x,\varepsilon )G^{{\rm cov}}(x,\varepsilon )-G^{{\rm cov}%
}(x,\varepsilon )G_\mu (x,-\varepsilon )  \nonumber \\
&&\ \ \ \ +H_\mu (x,\varepsilon )G^{{\rm cov}}(x,\varepsilon )+G^{{\rm cov}%
}(x,\varepsilon )H_\mu (x,-\varepsilon ));  \nonumber \\
&&
\end{eqnarray}
this is the formula (\ref{G-H formula}).

\vskip1cm {\large {\bf Appendix B }} \vskip1cm

In this appendix we give some details of the calculation within the
non-covariant point-splitting. Let us first derive the $\varepsilon $%
-expansion of the matrix functions $\Omega _{\pm }$. We have
\begin{eqnarray}
\Omega _{-} &=&\widetilde{\Omega }(x,x-\varepsilon )\Omega
^{+}(x,x-\varepsilon )={\rm e}^{-\varepsilon \cdot \widetilde{D}}{\rm e}%
^{\varepsilon \cdot D},  \nonumber \\
\Omega _{+} &=&\Omega ^{+}(x+\varepsilon ,x)\widetilde{\Omega }%
(x+\varepsilon ,x)={\rm e}^{-\varepsilon \cdot \stackrel{\leftarrow }{D}}%
{\rm e}^{\varepsilon \cdot \stackrel{\leftarrow }{\widetilde{D}}}.
\end{eqnarray}
where $\widetilde{D}=\partial +aA$ and $\stackrel{\leftarrow }{\widetilde{D}}%
=\stackrel{\leftarrow }{\partial }-aA$ are the non-covariant modification of
the covariant derivatives.
Using the Baker-Campbell-Hausdorff formula we get
\begin{eqnarray}
\Omega _{-} &=&\widetilde{\Omega }(x,x-\varepsilon )\Omega
^{+}(x,x-\varepsilon )={\rm e}^{-\varepsilon \cdot \widetilde{D}}{\rm e}%
^{\varepsilon \cdot D}
\nonumber \\
\ &=& {\rm e}^{-\varepsilon \cdot \widetilde{D}+ \varepsilon \cdot D
-\frac 12[\varepsilon \cdot \widetilde{D},\varepsilon \cdot D]
+\frac 1{12} [\varepsilon \cdot \widetilde{D},[\varepsilon \cdot \widetilde{D}%
,\varepsilon \cdot D]]
-\frac 1{12} [\varepsilon \cdot D,[\varepsilon \cdot D%
,\varepsilon \cdot \widetilde{D}]]
+{\cal O}(\varepsilon ^4)
}
\nonumber \\
\ &=&1+\varepsilon \cdot (D-\widetilde{D})
+\frac 12([\varepsilon \cdot
\widetilde{D},(\varepsilon \cdot \widetilde{D}-\varepsilon \cdot D)]
+(\varepsilon \cdot (D-\widetilde{D}))^2)
\nonumber \\
&&\ +\frac 16((\varepsilon \cdot (D-\widetilde{D}))^3
-\frac 14 \{ \varepsilon \cdot (D-\widetilde{D}),
[\varepsilon \cdot \widetilde{D},\varepsilon \cdot D] \}
+\frac 1{12} [ \varepsilon \cdot (D+\widetilde{D}),
[\varepsilon \cdot \widetilde{D},\varepsilon \cdot D] ]
\nonumber \\
&&+{\cal O}(\varepsilon ^4).
\end{eqnarray}
Using now $\varepsilon \cdot D-\varepsilon \cdot \widetilde{D}%
=(1-a)\varepsilon \cdot A$ , we rewrite this to the final form
\begin{eqnarray}
\Omega _{-} &=&1+(1-a)\varepsilon \cdot A+\frac 12(1-a)((1-a)(\varepsilon
\cdot A)^2-(\varepsilon \cdot \partial \varepsilon \cdot A)  \nonumber \\
&&\ +\frac 16(1-a)((\varepsilon \cdot \partial )^2\varepsilon \cdot
A+(1-a)^2(\varepsilon \cdot A)^3  \nonumber \\
&&\ +(2a-1)\varepsilon \cdot A\varepsilon \cdot \partial \varepsilon \cdot
A+(a-2)(\varepsilon \cdot \partial \varepsilon \cdot A)\varepsilon \cdot A)+%
{\cal O}(\varepsilon ^4).
\end{eqnarray}
If we realize, that $\Omega _{+}={\rm e}^{-\varepsilon \cdot \stackrel{%
\leftarrow }{D}}{\rm e}^{\varepsilon \cdot \stackrel{\leftarrow }{\widetilde{%
D}}}={\rm e}^{\varepsilon \cdot D}{\rm e}^{-\varepsilon \cdot \widetilde{D}%
}=({\rm e}^{\varepsilon \cdot \widetilde{D}}{\rm e}^{-\varepsilon \cdot
D})^{+}$, we can immediately obtain $\Omega _{+}$ by means of Hermitian
conjugation and replacing then $\varepsilon \rightarrow -\varepsilon $, i.e.
\begin{eqnarray}
\Omega _{+} &=&1+(1-a)\varepsilon \cdot A+\frac 12(1-a)((1-a)(\varepsilon
\cdot A)^2+(\varepsilon \cdot \partial \varepsilon \cdot A)  \nonumber \\
&&\ \ +\frac 16(1-a)((\varepsilon \cdot \partial )^2\varepsilon \cdot
A+(1-a)^2(\varepsilon \cdot A)^3  \nonumber \\
&&\ \ -(2a-1)(\varepsilon \cdot \partial \varepsilon \cdot A)\varepsilon
\cdot A-(a-2)\varepsilon \cdot A\varepsilon \cdot \partial \varepsilon \cdot
A)+{\cal O}(\varepsilon ^4).
\end{eqnarray}
It is now straightforward to get the formula
\begin{eqnarray}
\Omega _{-}\Omega _{+} &=&1+2(1-a)\varepsilon \cdot A+2(1-a)^2(\varepsilon
\cdot A)^2  \nonumber \\
&&\ +\frac 43(1-a)^3(\varepsilon \cdot A)^3+\frac 13(1-a)(\varepsilon \cdot
\partial )^2\varepsilon \cdot A  \nonumber \\
&&\ +\frac 13(1-a)(2-a)[\varepsilon \cdot A,\varepsilon \cdot \partial
\varepsilon \cdot A]+{\cal O}(\varepsilon ^4),
\end{eqnarray}
which was used in the main text. We need also the expansion of the trace $%
{\rm Tr}_C(\Omega _{+}\Omega _{-}-1)$ to the fourth order in $\varepsilon $.
We have, using the operator expression for $\Omega _{\pm }$, for the fourth
term of the $\varepsilon $ expansion
\begin{eqnarray}
{\rm Tr}_C(\Omega _{+}\Omega _{-})^{(4)} &=&\frac 12{\rm Tr}_C(\Omega
_{+}\Omega _{-}+\Omega _{-}\Omega _{+})^{(4)}  \nonumber \\
\ &=&\frac 12{\rm Tr}_C(\frac 43(\varepsilon \cdot D)^4+\frac 43(\varepsilon
\cdot \widetilde{D})^4-\frac 43\{\varepsilon \cdot D,(\varepsilon \cdot
\widetilde{D})^3\}-\frac 43\{\varepsilon \cdot \widetilde{D},(\varepsilon
\cdot D)^3\}  \nonumber \\
&&\ +\{\varepsilon \cdot D,\{\varepsilon \cdot D,(\varepsilon \cdot
\widetilde{D})^2\}\}+\{\varepsilon \cdot \widetilde{D},\{\varepsilon \cdot
\widetilde{D},(\varepsilon \cdot D)^2\}\}  \nonumber \\
&&\ -\frac 13\{\varepsilon \cdot D,\{\varepsilon \cdot D,\{\varepsilon \cdot
D,\varepsilon \cdot \widetilde{D}\}\}\}  \nonumber \\
&&-\frac 13\{\varepsilon \cdot \widetilde{D},\{\varepsilon \cdot \widetilde{D%
},\{\varepsilon \cdot \widetilde{D},\varepsilon \cdot D\}\}\}).
\end{eqnarray}
After some algebra we get
\begin{eqnarray}
{\rm Tr}_C(\Omega _{+}\Omega _{-})^{(4)} &=&\frac 16{\rm Tr}_C(4(\varepsilon
\cdot (D-\widetilde{D}))^4+\{\varepsilon \cdot (D-\widetilde{D}%
),[\varepsilon \cdot (D+\widetilde{D}),[\varepsilon \cdot \widetilde{D}%
,\varepsilon \cdot (D-\widetilde{D})]]\})  \nonumber \\
\ &=&\frac 23{\rm Tr}_C((1-a)^4(\varepsilon \cdot A)^4+(1-a)^2\varepsilon
\cdot A(\varepsilon \cdot \partial )^2\varepsilon \cdot A).
\end{eqnarray}
Let us now work out the expression $\Omega _{+}[\stackrel{\rightarrow }{D}%
_\nu ,\Omega _{-}]-[\Omega _{+},\stackrel{\leftarrow }{D}_\nu ]\Omega _{-}$.
We have
\begin{eqnarray}
\Omega _{+}[\stackrel{\rightarrow }{D}_\nu ,\Omega _{-}]-[\Omega _{+},%
\stackrel{\leftarrow }{D}_\nu ]\Omega _{-} &=&\Omega _{+}[\stackrel{%
\rightarrow }{D}_\nu ,{\rm e}^{-\varepsilon \cdot \widetilde{D}}{\rm e}%
^{\varepsilon \cdot D}]-[{\rm e}^{-\varepsilon \cdot \stackrel{\leftarrow }{D%
}}{\rm e}^{\varepsilon \cdot \stackrel{\leftarrow }{\widetilde{D}}},%
\stackrel{\leftarrow }{D}_\nu ]\Omega _{-}  \nonumber \\
\ &=&\Omega _{+}[\stackrel{\rightarrow }{D}_\nu ,{\rm e}^{-\varepsilon \cdot
\widetilde{D}}]{\rm e}^{\varepsilon \cdot \widetilde{D}}{\rm e}%
^{-\varepsilon \cdot \widetilde{D}}{\rm e}^{\varepsilon \cdot D}+\Omega _{+}%
{\rm e}^{-\varepsilon \cdot \widetilde{D}}{\rm e}^{\varepsilon \cdot D}{\rm e%
}^{-\varepsilon \cdot D}[\stackrel{\rightarrow }{D}_\nu ,{\rm e}%
^{\varepsilon \cdot D}]-  \nonumber \\
&&\ -{\rm e}^{-\varepsilon \cdot \stackrel{\leftarrow }{D}}{\rm e}%
^{\varepsilon \cdot \stackrel{\leftarrow }{\widetilde{D}}}{\rm e}%
^{-\varepsilon \cdot \stackrel{\leftarrow }{\widetilde{D}}}[{\rm e}%
^{\varepsilon \cdot \stackrel{\leftarrow }{\widetilde{D}}},\stackrel{%
\leftarrow }{D}_\nu ]\Omega _{-}-[{\rm e}^{-\varepsilon \cdot \stackrel{%
\leftarrow }{D}},\stackrel{\leftarrow }{D}_\nu ]{\rm e}^{\varepsilon \cdot
\stackrel{\leftarrow }{D}}{\rm e}^{-\varepsilon \cdot \stackrel{\leftarrow }{%
D}}{\rm e}^{\varepsilon \cdot \stackrel{\leftarrow }{\widetilde{D}}}\Omega
_{-}  \nonumber \\
\ &=&-\Omega _{+}\widetilde{G}_\nu (x,\varepsilon )\Omega _{-}+\Omega
_{+}\Omega _{-}G_\nu (x,\varepsilon )  \nonumber \\
&&\ -\Omega _{+}\widetilde{G}_\nu (x,-\varepsilon )\Omega _{-}+G_\nu
(x,-\varepsilon )\Omega _{+}\Omega _{-},
\end{eqnarray}
where
\begin{equation}
\widetilde{G}_\nu (x,\varepsilon )=[{\rm e}^{-\varepsilon \cdot \widetilde{D}%
}\stackrel{\rightarrow }{,D}_\nu ]{\rm e}^{\varepsilon \cdot \widetilde{D}%
}=\sum_{n=0}^\infty \frac{(-1)^n}{(n+1)!}[\varepsilon \cdot \widetilde{D}%
,[\varepsilon \cdot \widetilde{D},[\ldots ,[\varepsilon \cdot \widetilde{D}%
,D_\nu ]\ldots ]]]
\end{equation}
is a non-covariant modification of the function $G_\nu (x,\varepsilon )$
introduced above. For the calculation of the counterterm $\Delta \widetilde{%
\theta }_{\mu \nu }$ we have derived
\begin{eqnarray}
\Delta \widetilde{\theta }_{\mu \nu }^{{\rm reg}} &=&-\frac 12{\rm Tr}S^{%
{\rm cov}}\gamma _\mu (\Omega _{+}[\stackrel{\rightarrow }{D}_\nu ,\Omega
_{-}]-[\Omega _{+},\stackrel{\leftarrow }{D}_\nu ]\Omega _{-}  \nonumber \\
&&\ \ \ \ \ +(\Omega _{+}\Omega _{-}-1)\Gamma _\nu (x,\varepsilon )+\Gamma
_\nu (x,-\varepsilon )(\Omega _{+}\Omega _{-}-1))  \nonumber \\
&&\ \ \ \ \ \ \ +\frac 12{\rm Tr}(\partial _\nu ^\varepsilon S^{{\rm cov}%
}\gamma _\mu )(\Omega _{+}\Omega _{-}-1).
\end{eqnarray}
Remembering the formula for the ${\rm Tr}_DS^{{\rm cov}}\gamma _\mu $ and
noting that
\begin{equation}
\Omega _{+}[\stackrel{\rightarrow }{D}_\nu ,\Omega _{-}]-[\Omega _{+},%
\stackrel{\leftarrow }{D}_\nu ]\Omega _{-}+(\Omega _{+}\Omega _{-}-1)\Gamma
_\nu (x,\varepsilon )+\Gamma _\nu (x,-\varepsilon )(\Omega _{+}\Omega
_{-}-1)={\cal O}(\varepsilon ^2)  \label{expr}
\end{equation}
it is easily seen that we need only the trace over the color indices of the
expression (\ref{expr}) expanded to the order ${\cal O}(\varepsilon ^3)$. We
have then up to the terms which do not contribute to the final expression
after the symmetrization over the direction of $\varepsilon $
\begin{eqnarray}
&&\ \ \ \ \ \ {\rm Tr}_C((\Omega _{+}\Omega _{-}-1)(\Gamma _\nu
(x,\varepsilon )+\Gamma _\nu (x,-\varepsilon ))+\Omega _{+}[\stackrel{%
\rightarrow }{D}_\nu ,\Omega _{-}]-[\Omega _{+},\stackrel{\leftarrow }{D}%
_\nu ]\Omega _{-})  \nonumber \\
\ &=&-\frac 23(1-a){\rm Tr}_C((2-3a)(\varepsilon \cdot A)[D_\nu
,(\varepsilon \cdot \partial )(\varepsilon \cdot A)]+(\varepsilon \cdot
A)(\varepsilon \cdot \partial )^2A_\nu )+{\cal O}(\varepsilon ^4).  \nonumber
\\
&&  \label{factor1}
\end{eqnarray}
Putting all the ingredients together we get
\begin{eqnarray}
\Delta \widetilde{\theta }_{\mu \nu }^{{\rm reg}}
&=&\frac{1-a}{8\pi ^2}{\rm Tr}_C
\biggl(-\frac 1{36}((2-3a)(A_\mu \partial _\nu \partial \cdot A+A_\alpha
\partial _\mu \partial _\nu A_\alpha +A\cdot \partial \partial _\nu A_\mu
\nonumber \\
&&\ \ \ +A_\mu [A_\nu ,\partial \cdot A]+A_\alpha [A_\nu ,\partial _\mu
A_\alpha ]+A_\alpha [A_\nu ,\partial _\alpha A_\mu ])  \nonumber \\
&&\ \ \ +A_\mu \partial ^2A_\nu +2A\cdot \partial \partial _\mu A_\nu )
\nonumber \\
&&\ \ \ -\frac{(1-a)^3}{72}(\delta _{\mu \nu }(2A^4+A_\alpha A_\beta
A_\alpha A_\beta )-4(\{A_\mu ,A_\nu \}A^2+A_\mu A_\alpha A_\nu A_\alpha ))
\nonumber \\
&&\ \ \ -\frac{(1-a)}{72}(\delta _{\mu \nu }(2A\cdot \partial \partial \cdot
A+A_\alpha \partial ^2A_\alpha )-2(A_\mu \partial _\nu +A_\nu \partial _\mu
)\partial \cdot A  \nonumber \\
&&\ \ \ -2A\cdot \partial (\partial _\nu A_\mu +\partial _\mu A_\nu )-A_\mu
\partial ^2A_\nu -A_\nu \partial ^2A_\mu -2A_\alpha \partial _\mu \partial
_\nu A_\alpha )  \nonumber \\
&&\ \ \ +\frac{1-a}6\left(\delta _{\mu \nu }A^2
\left(2m^2-\frac 1{|\varepsilon|^2}\right)
-2A_\mu A_\nu \left(m^2-\frac 2{|\varepsilon |^2}\right)\right)
\nonumber \\
&&\ \ \ -\frac 19(\delta _{\mu \nu }A_\alpha [D_\beta ,F_{\beta \alpha
}]+A_\mu [D_\alpha ,F_{\alpha \nu }]-4A_\nu [D_\alpha ,F_{\alpha \mu }]
\nonumber  \\
&&\ \ -A_\alpha [D_\alpha ,F_{\mu \nu }]
-A_\alpha [D_\nu ,F_{\mu \alpha }])\biggr).
\label{deltatheta}
\end{eqnarray}
Using the following identities
\begin{eqnarray}
{\rm Tr}_C(A_\mu [D_\alpha ,F_{\alpha \nu }]+A_\nu [D_\alpha ,F_{\alpha \mu
}]) &=&{\rm Tr}_C(A_\mu \partial ^2A_\nu +A_\nu \partial ^2A_\mu -(A_\mu
\partial _\nu +A_\nu \partial _\mu )\partial \cdot A  \nonumber \\
&&\ +2A_\alpha [\partial _\alpha A_\nu ,A_\mu ]+2A_\alpha [\partial _\alpha
A_\mu ,A_\nu ]  \nonumber \\
&&\ -A_\alpha [\partial _\nu A_\alpha ,A_\mu ]-A_\alpha [\partial _\mu
A_\alpha ,A_\nu ]  \nonumber \\
&&\ +2\{A_\mu ,A_\nu \}A^2-4A_\alpha A_\mu A_\alpha A_\nu )
\end{eqnarray}

\begin{eqnarray}
{\rm Tr}_C(A_\alpha [D_\nu ,F_{\mu \alpha }]+A_\alpha [D_\mu ,F_{\nu \alpha
}]) &=&{\rm Tr}_C(2A_\alpha \partial _\mu \partial _\nu A_\alpha -A\cdot
\partial (\partial _\nu A_\mu +\partial _\mu A_\nu )  \nonumber \\
&&\ \ \ -2A_\alpha [\partial _\nu A_\alpha ,A_\mu ]-2A_\alpha [\partial _\mu
A_\alpha ,A_\nu ]  \nonumber \\
&&\ \ \ +A_\alpha [\partial _\alpha A_\nu ,A_\mu ]+A_\alpha [\partial
_\alpha A_\mu ,A_\nu ]  \nonumber \\
&&\ \ \ +2\{A_\mu ,A_\nu \}A^2-4A_\alpha A_\mu A_\alpha A_\nu )
\end{eqnarray}
\begin{eqnarray}
{\rm Tr}_C(A_\mu [D_\alpha ,F_{\alpha \nu }]-A_\nu [D_\alpha ,F_{\alpha \mu
}]) &=&{\rm Tr}_C(A_\mu \partial ^2A_\nu -A_\nu \partial ^2A_\mu -2A\cdot
\partial [A_\mu ,A_\nu ]  \nonumber \\
&&\ \ \ -(A_\mu \partial _\nu -A_\nu \partial _\mu +2[A_\mu ,A_\nu
])\partial \cdot A  \nonumber \\
&&\ \ \ -A_\mu [A_\alpha ,\partial _\nu A_\alpha ]+A_\nu [A_\alpha ,\partial
_\mu A_\alpha ])
\end{eqnarray}
we get
\begin{eqnarray}
\Delta \widetilde{\theta }_{\mu \nu }^{{\rm reg}}
&=&\frac{(1-a)^2}{48\pi ^2}{\rm Tr}_C
\left(\delta _{\mu \nu }A^2\left(2m^2-\frac 1{|\varepsilon|^2}\right)
-2A_\mu A_\nu \left(m^2-\frac 2{|\varepsilon |^2}\right)\right)
\nonumber \\
&&\ \ \ \ -\frac{(1-a)}{576\pi ^2}{\rm Tr}_C(8\delta _{\mu \nu }A_\alpha
[D_\beta ,F_{\beta \alpha }]-11(A_\mu [D_\alpha ,F_{\alpha \nu }]+A_\nu
[D_\alpha ,F_{\alpha \mu }])  \nonumber \\
&&\ \ \ \ -5(A_\alpha [D_\nu ,F_{\mu \alpha }]+A_\alpha [D_\mu ,F_{\nu
\alpha }])  \nonumber \\
&&\ \ \ \ +(1-a)(\delta _{\mu \nu }(2A\cdot \partial \partial \cdot
A+A_\alpha \partial ^2A_\alpha )+(A_\mu \partial _\nu +A_\nu \partial _\mu
)\partial \cdot A  \nonumber \\
&&\ \ \ \ +A\cdot \partial (\partial _\nu A_\mu +\partial _\mu A_\nu
)+4A_\alpha \partial _\mu \partial _\nu A_\alpha -A_\mu \partial ^2A_\nu
-A_\nu \partial ^2A_\mu  \nonumber \\
&&\ \ \ \ +3([A_\alpha ,A_\mu ](\partial _\nu A_\alpha +\partial _\alpha
A_\nu )+[A_\alpha ,A_\nu ](\partial _\mu A_\alpha +\partial _\alpha A_\mu )))
\nonumber \\
&&\ \ \ \ +(1-a)^2(\delta _{\mu \nu }(2A^4+A_\alpha A_\beta A_\alpha A_\beta
)-4(\{A_\mu ,A_\nu \}A^2+A_\mu A_\alpha A_\nu A_\alpha )))  \nonumber \\
&&\ \ \ \ -\frac{(1-a)}{192\pi ^2}{\rm Tr}_C((a+6)(A_\mu [D_\alpha
,F_{\alpha \nu }]-A_\nu [D_\alpha ,F_{\alpha \mu }])+(a-4)A\cdot \partial
F_{\mu \nu }  \nonumber \\
&&\ \ \ \ +(1-a)(A_\mu \partial ^2A_\nu -A_\nu \partial ^2A_\mu -2A\cdot
\partial [A_\mu ,A_\nu ]));
\end{eqnarray}
this is the formula (\ref{counterterm theta}).

\end{document}